\documentclass{pasa}%
\usepackage{lscape, morefloats, graphicx}
\title[Local stellar kinematics from RAVE data - VII.]{Local Stellar Kinematics from RAVE data - VII. Metallicity Gradients from Red Clump Stars}
\author[\"Onal Tas et al.]{\"O. \"Onal Ta\c s$^1$\thanks{ozgecan.onal@istanbul.edu.tr}, S. Bilir$^2$, G.~M. Seabroke$^3$, S. Karaali$^2$, S. Ak$^2$, T. Ak$^2$, and Z.~F. Bostanc\i$^2$\\
     \affil{$^1$Istanbul University, Graduate School of Science and Engineering, Department of Astronomy and Space Sciences, 34116, Beyaz\i t, Istanbul, Turkey}
     \affil{$^2$Istanbul University, Faculty of Science, Department of Astronomy and Space Sciences, 34119, Beyaz\i t, Istanbul, Turkey}
     \affil{$^3$University College London, Mullard Space Science Laboratory, Holmbury St Mary, Dorking, RH5 6NT, United Kingdom}
}
\jid{PASA}
\doi{10.1017/pas.\the\year.xxx}
\jyear{\the\year}
\usepackage{natbib}
\usepackage{graphicx}
\graphicspath{%
    {converted_graphics/}
    {/}
}
\begin{document}
\begin{abstract}
We investigate the Milky Way Galaxy's radial and vertical metallicity gradients using a sample of 47,406 red clump stars from the RAdial Velocity Experiment (RAVE) Data Release (DR) 4. This sample is more than twice the size of the largest sample in the literature investigating radial and vertical metallicity gradients. The absolute magnitude of \citet{Groenewegen08} is used to determine distances to our sample stars. The resulting distances agree with the RAVE DR4 distances \citep{Binney14} of the same stars. Our photometric method also provides distances to 6185 stars that are not assigned a distance in RAVE DR4. The metallicity gradients are calculated with their current orbital positions ($R_{gc}$ and $Z$) and with their orbital properties (mean Galactocentric distance, $R_{m}$ and $z_{max}$), as a function of the distance to the Galactic plane: d[Fe/H]/d$R_{gc}=$-$0.047\pm0.003$ dex kpc$^{-1}$ for $0\leq |Z|\leq0.5$ kpc and d[Fe/H]/d$R_m=$-$0.025\pm0.002$ dex kpc$^{-1}$ for $0\leq z_{max}\leq0.5$ kpc.  This reaffirms the radial metallicity gradient in the thin disc but highlights that gradients are sensitive to the selection effects caused by the difference between $R_{gc}$ and $R_{m}$. The radial gradient is flat in the distance interval 0.5-1 kpc from the plane and then becomes positive greater than 1 kpc from the plane. The radial metallicity gradients are also eccentricity dependent. We showed that d[Fe/H]/d$R_m=$-$0.089\pm0.010$, -$0.073\pm0.007$, -$0.053\pm0.004$ and -$0.044\pm0.002$ dex kpc$^{-1}$ for $e_p\leq0.05$, $e_p\leq0.07$, $e_p\leq0.10$ and $e_p\leq0.20$ sub-samples, respectively, in the distance interval $0\leq z_{max}\leq0.5$ kpc. Similar trend is found for vertical metallicity gradients. Both the radial and vertical metallicity gradients are found to become shallower as the eccentricity of the sample increases. These findings can be used to constrain different formation scenarios of the thick and thin discs.

\end{abstract}
\maketitle%

\section{Introduction}
How galaxies formed in general and how our Galaxy formed specifically remain as unsolved problem in astrophysics. In order to understand the formation mechanisms and put constraints on simulations, Galactic archaeology relies on chemo-dynamical information of various tracer objects with different ages and chemical compositions with large baselines in the Galaxy. While overlapping properties make it difficult to disentangle the different components in the Galaxy, there are clear chemical and kinematical signatures in stars of the Solar neighborhood which change with increasing distances both in radial and vertical directions. 
  
Sensitive metallicity gradients require tracer objects with well-measured distances and metallicities, which are obtained with either spectroscopic or photometric methods.  Tracers include main-sequence stars, red clump (RC) stars, open or globular clusters, Cepheid variables, planetary nebulae, O-B type stars and HII regions. Various distance scales are used to determine radial and vertical metallicity gradients in the literature. In radial gradient calculations, if the tracer objects are distant, then the current distance from the Galactic centre is used ($R_{gc}$). If proper motions, radial velocities and distances of tracer objects are known, even though the objects are currently close to the Sun, then the mean Galactocentric distances of the tracer objects calculated from Galactic orbital parameters are used, under the assumption of that the metallicities are constant on the orbital integration timescale ($R_m$). Another distance scale presented to the literature is the guiding radius which is the distance of the tracer from the Galactic centre, if it is following its guiding centre, assumed to be a circular orbit ($R_{g}$). Also, current vertical distances and maximum vertical distances from the Galactic plane of the objects are generally considered in vertical metallicity gradient calculations. 

Radial and vertical metallicity gradient studies that appeared in the literature almost in the last 15 years are listed in Tables 1 and 2, respectively. The organization of the metallicity gradient information is based on the distance indicators mentioned in the previous paragraph. According to Table 1, radial metallicity gradients tend to have steeper values near the Galactic plane and they flatten and even have positive values as the distance of the tracers from the Galactic plane increases. This behavior holds for different tracer objects, as well. Radial metallicity gradients of young objects show steeper values than the ones calculated from older objects. Similar results are also found in Table 1 for $R_m$. Vertical metallicity gradients are found to be steeper than their radial counterparts. And besides, absolute vertical metallicity gradients found for the Galactic disc ($z<5$ kpc) give larger values than the ones found for the Galactic halo ($5<z<10$ kpc; see Table 2). This is also valid for objects with distances calculated from different methods.

The {\it Hipparcos} \citep{ESA97} era calculated accurate distance estimations for stars in the Solar neighborhood. High-resolution spectroscopic follow-up observations with ground-based telescopes allowed astronomers to determine more precise metallicity gradients for the small {\it Hipparcos} volume in the Solar neighborhood \citep[i.e.][]{Nordstrom04}. To increase this limited volume, astronomers initiated several photometric and spectroscopic sky surveys, such as the RAdial Velocity Experiment \citep[RAVE;][]{Steinmetz06}, the Sloan Extension for Galactic Understanding and Exploration \citep[SEGUE;][]{Yanny09}, the APO-Galactic Evolution Experiment \citep[APOGEE;][]{Majewski10}, the Large sky Area Multi Object fiber Spectroscopic Telescope \citep[LAMOST;][]{Zhao12}, the GALactic Archaeology with HERMES \citep[GALAH;][]{Zucker12} and the Gaia-ESO Survey \citep[GES;][]{Gilmore12}. These surveys are designed to reveal the structure and test the formation mechanisms of the Milky Way's disc as well as to decode its evolution via metallicity, kinematics and dynamics of its stars.

According to stellar evolution models, a star within a mass range of 0.8 to 2.2 $M_{\odot}$ could generate a helium (He) core mass of $\sim0.45M_{\odot}$ after the hydrogen burning reactions ceased \citep{Girardi99}. At this point, as stated by \citet{Carretta99}, when a star begins to evolve to the red giant phase, its surrounding envelope starts to interact with inner regions of the star, thus materials inside and in the outer surface of the star start mixing via convection processes. In a red giant, gravitational collapse of the star is supported via pressure of degenerate electrons, until the He core reaches 0.45 $M_{\odot}$. Then a series of He-flashes \citep{Thomas67} occurs and breaks the degeneracy. Right after the removal of degeneracy, luminosity of the star suddenly drops and stable He burning reactions begin. Star's luminosity remains almost constant. At this stage, if a star is metal rich then it becomes an RC star, or if it is not metal rich then it becomes a member of the blue horizontal branch stars \citep{Iben91}. Since the main-sequence mass of the RC stars are between 0.9 and 2.25 $M_{\odot}$ according to theory \citep{Eggleton68}, their main-sequence life times are between 13 to 1.3 Gyrs, respectively. Theory also predicts that the horizontal branch life times of stars is roughly 10\% of the main-sequence life times. This is the reason why these stars form a clump structure in Hertzsprung-Russell diagrams. There has been much debate on the nature of the RC stars and their use as a standard candle since their discovery. \citet {Cannon70} suggested that RC stars could be used for distance determination as standard candles. There are several studies which show that the absolute magnitude of the RC stars in different parts of the electromagnetic spectrum is always the same \citep {Paczynski98, Keenan99, Alves00, Udalski00, Groenewegen08, Laney12, Bilir13a, Bilir13b, Karaali13, Yaz13}: absolute magnitudes of the RC stars are in the near infrared $K_s$ band, $M_{Ks}=$-$1.54\pm 0.04$ mag \citep {Groenewegen08}, and in $I$ band $M_I=$-$0.23\pm 0.03$ mag \citep {Paczynski98}. RC stars are common objects in the Solar neighborhood and their intrinsic brightness probes large distances, which enables us to investigate chemical gradients both in radial and vertical directions.

Rather than using spectroscopic parallaxes, we preferred photometric parallax method to calculate distances using the aforementioned well-known constant infrared absolute magnitude in the $K_s$ band. Metallicity gradients are calculated for the current orbital positions of the stars as well as for their calculated complete orbit parameters. In metallicity gradient calculations we used the mean Galactocentric distance of stellar orbits ($R_{m}$) instead of guiding radius ($R_{g}$). One other aspect of the study is that the dynamical properties of each star are calculated with {\em galpy} library of \citet {Bovy15} using test particle integration for {\it MWPotential2014} potential.

In this paper, we present results on the metallicity gradients of  the RC stars using data from RAVE DR4. In \S 2, we present the selection criteria of the RC stars and their distance, kinematic and dynamic calculations. In \S 3, we give the results on overall metallicity gradients obtained using various distance scales both in radial and vertical directions in the Galaxy. Also, a sub-sample separation with planar and vertical eccentricities are carried out and implications on radial migration of the RC stars are mentioned. We discussed the results on the RC stars in \S 4.
\\
\begin{landscape}
\textwidth = 650pt
\begin{table*}
\setlength{\tabcolsep}{2.5pt}
\centering
{\tiny
\caption{Radial metallicity gradients appeared in the literature. Distances ($R_{gc}$, $Z$, $R_{m}$, $z_{max}$, $R_{g}$) in kpc, age ($\tau$) in Gyr.}
\begin{tabular}{lcccccccr}
\hline
~~~~~~~~~~~~~~Author & Tracer Object & d[Fe/H]/d$R_{gc}$ & Remark & d[Fe/H]/d$R_{m}$ & Remark & d[Fe/H]/d$R_{g}$ & Remark & Bibliographic Code \\
      & &(dex~kpc$^{-1}$)& &(dex~kpc$^{-1}$)& &(dex~kpc$^{-1}$)& & \\
\hline
\citet{Jacobson16}      & OC                       & -$0.100\pm0.020$ & $6<R_{gc}<12$&$-$ &$-$ &$-$ &$-$ &2016A\&A...591A..37J\\
\citet{Cunha16}         & OC/RG                    & -$0.035\pm0.007$ & $6<R_{gc}<25$&$-$ &$-$ &$-$ &$-$    &arXiv:1601.03099\\
\citet{Netopil16}       & OC                       & -$0.086\pm0.009$ & $5<R_{gc}<12$&$-$ &$-$ &$-$ &$-$    &2016A\&A...585A.150N\\
                        &                          & -$0.082\pm0.013$ & $1<\tau<2.5$ &$-$ &$-$ &$-$ &$-$    &2016A\&A...585A.150N\\
\citet{Plevne15}        & FG MSs                   & $-$ & $-$ & -$0.083\pm0.030$& $0<z_{max}\leq0.5$ ([Fe/H]) & -$0.083\pm0.030$ & $0<z_{max}\leq0.5$ ([Fe/H]) & 2015PASA...32...43P \\
                        &                          & $-$ & $-$ & -$0.048\pm0.037$& $0.5<z_{max}\leq0.8$ ([Fe/H]) & -$0.065\pm0.039$ & $0.5<z_{max}\leq0.8$ ([Fe/H]) &  2015PASA...32...43P\\  
                        &                          & $-$ & $-$ & -$0.063\pm0.011$& $0<z_{max}\leq0.5$ (M/H]) & -$0.062\pm0.018$ & $0<z_{max}\leq0.5$ ([M/H]) & 2015PASA...32...43P \\
                        &                          & $-$ & $-$ & -$0.028\pm0.057$& $0.5<z_{max}\leq0.8$ ([M/H]) & -$0.055\pm0.045$ & $0.5<z_{max}\leq0.8$ ([M/H]) & 2015PASA...32...43P \\    
\citet{Huang15}         & RCs  			   & -$0.082\pm0.003$ & $|Z|\leq0.1$ &$-$ &$-$ &$-$ &$-$ & 2015RAA....15.1240H\\
	                &   			   & -$0.072\pm0.004$ & $0.1<|Z|\leq0.3$ &$-$ &$-$ &$-$ &$-$ & 2015RAA....15.1240H\\
	                &   			   & -$0.052\pm0.005$ & $0.3<|Z|\leq0.5$ &$-$ &$-$ &$-$ &$-$ & 2015RAA....15.1240H\\
	                &   			   & -$0.041\pm0.005$ & $0.5<|Z|\leq0.7$ &$-$ &$-$ &$-$ &$-$ & 2015RAA....15.1240H\\
	                &   			   & -$0.028\pm0.005$ & $0.7<|Z|\leq0.9$ &$-$ &$-$ &$-$ &$-$ & 2015RAA....15.1240H\\
	                &   			   & -$0.020\pm0.007$ & $0.9<|Z|\leq1.1$ &$-$ &$-$ &$-$ &$-$ & 2015RAA....15.1240H\\
\citet{Xiang15}         & Turnoff stars		   & -$0.100\pm0.003$ & $|Z|\leq0.1; 2<\tau<16$ &$-$ &$-$ &$-$ &$-$ &  2015RAA....15.1209X\\
                        & 		           & -$0.050\pm0.002$ & $0.4<|Z|\leq0.6; 2<\tau<16$ &$-$ &$-$ &$-$ &$-$ &  2015RAA....15.1209X\\
                        & 		           & -$0.010\pm0.002$ & $0.9<|Z|\leq1.1; 2<\tau<16$ &$-$ &$-$ &$-$ &$-$ &  2015RAA....15.1209X\\
\citet{Recio14}         & FGK MSs 	      	   & -$0.058\pm0.008$ & Thin disc ([M/H])  &$-$ &$-$ &$-$ &$-$ & 2014A\&A...567A...5R\\
			&   			   & +$0.006\pm0.008$ & Thick disc ([M/H]) &$-$ &$-$ &$-$ &$-$ & 2014A\&A...567A...5R\\		
\citet{Mikolaitis14}    & FGK MSs 	           & -$0.044\pm0.009$ & Thin disc (main)   &$-$ &$-$ &$-$ &$-$ & 2014A\&A...572A..33M\\
    			&   			   & -$0.028\pm0.018$ & Thin disc (clean)  &$-$ &$-$ &$-$ &$-$ & 2014A\&A...572A..33M\\
    			&   			   & +$0.008\pm0.007$ & Thick disc (main)  &$-$ &$-$ &$-$ &$-$ & 2014A\&A...572A..33M\\
    			&   			   & +$0.008\pm0.007$ & Thick disc (clean) &$-$ &$-$ &$-$ &$-$ & 2014A\&A...572A..33M\\
\citet{Genovali14}      & Cepheids                 & -$0.021\pm0.029$ & $5<R_{gc}<19$ & $-$ & $-$ & $-$ & $-$ & 2014A\&A...566A..37G \\
\citet{Andreuzzi14}     & OC                       & -$0.04$          & $6<R_{gc}<22$ & $-$ & $-$ & $-$ & $-$ & 2011MNRAS.412.1265A \\
\citet{Boeche14}        & RCs                      & $-$&$-$ &$-$ & $-$ &-$0.054\pm 0.004$ & $  0<|Z|\leq0.4$ & 2014A\&A...568A..71B\\
                        &                          & $-$&$-$ &$-$ & $-$ &-$0.039\pm 0.004$ & $0.4<|Z|\leq0.8$ & 2014A\&A...568A..71B\\
                        &                          & $-$&$-$ &$-$ & $-$ &-$0.011\pm 0.008$ & $0.8<|Z|\leq1.2$ & 2014A\&A...568A..71B\\
                        &                          & $-$&$-$ &$-$ & $-$ &+$0.047\pm 0.018$ & $1.2<|Z|\leq2.0$ & 2014A\&A...568A..71B\\
\citet{Hayden14}        & RG                       & -$0.090\pm 0.002$ & $0.00<|Z|<0.25$; $0<R_{gc}<15$ & $-$& $-$& $-$& $-$& 2014AJ....147..116H\\
                        &                          & -$0.076\pm 0.003$ & $0.25<|Z|<0.50$; $0<R_{gc}<15$ & $-$& $-$& $-$& $-$& 2014AJ....147..116H\\
  	                &                          & -$0.057\pm 0.003$ & $0.50<|Z|<1.00$; $0<R_{gc}<15$ & $-$& $-$& $-$& $-$& 2014AJ....147..116H\\
  	                &                          & -$0.030\pm 0.006$ & $1.00<|Z|<2.00$; $0<R_{gc}<15$ & $-$& $-$& $-$& $-$& 2014AJ....147..116H\\
\citet{Frinchaboy13}    & OC                       & -$0.090\pm 0.030$ & $7.9\leq R_{gc} \leq 14.5$ & $-$ & $-$ & $-$ & $-$ & 2013ApJ...777L...1F\\
	                &                          & -$0.200\pm 0.080$ & $7.9\leq R_{gc} \leq 10.0$ & $-$ &$-$ & $-$& $-$& 2013ApJ...777L...1F\\
	                &                          & -$0.020\pm 0.090$ & $10<R_{gc} \leq 14.5$ & $-$ &$-$ & $-$& $-$& 2013ApJ...777L...1F\\
\citet{Boeche13}        & FG MSs                   & -$0.059\pm 0.002$ & $4.5<R_{gc}<9.5$ & $-$& $-$& -$0.065\pm0.003$ & $0<z_{max}\leq0.4$ & 2013A\&A...559A..59B\\
	                &                          & $-$ & $-$ & $-$&$-$ & -$0.059\pm0.006$ & $0.4<z_{max}\leq0.8$ & 2013A\&A...559A..59B\\
	                &                          & $-$ & $-$ & $-$&$-$ & +$0.006\pm0.015$ & $z_{max}>0.8$ & 2013A\&A...559A..59B\\
\citet{Carrell12}       & FGK MSs                  & +$0.015\pm 0.005$ & $0.5<|Z|<1.0$, Isoc. method   &$-$ &$-$ &$-$ &$-$ & 2012AJ....144..185C\\
	                &                          & +$0.017\pm 0.005$ & $0.5<|Z|<1.0$, Photo. method &$-$ &$-$ &$-$ &$-$ & 2012AJ....144..185C\\			
\citet{Bilir12}         & RCs                      & $-$& $-$ &-$0.041\pm0.003$ & $TD/D\leq 0.1$ &$-$ &$-$ & 2012MNRAS.421.3362B\\
	                &                          & $-$& $-$ &-$0.041\pm0.007$ & $e_v\leq 0.07$ &$-$ &$-$ & 2012MNRAS.421.3362B\\
	                &                          & $-$& $-$ &-$0.025\pm0.040$ & $e_v\leq 0.12$ &$-$ &$-$ & 2012MNRAS.421.3362B\\
	                &                          & $-$& $-$ &+$0.022\pm0.006$ & $e_v>0.25$ &$-$ &$-$ & 2012MNRAS.421.3362B\\
\citet{Cokkun12}        & FG MSs                   & $-$& $-$ &-$0.043\pm0.005$ & F stars, $TD/D\leq0.1  $&$-$ &$-$ & 2012MNRAS.419.2844C\\
                        &                          & $-$& $-$ &-$0.033\pm0.007$ & G stars, $TD/D\leq0.1  $&$-$ &$-$ & 2012MNRAS.419.2844C\\
                        &                          & $-$& $-$ &-$0.051\pm0.005$ & F stars, $e_v\leq 0.04 $&$-$ &$-$ & 2012MNRAS.419.2844C\\
                        &                          & $-$& $-$ &-$0.020\pm0.006$ & G stars, $e_v\leq 0.04 $&$-$ &$-$ & 2012MNRAS.419.2844C\\
\citet{Cheng12}         & Turnoff stars            & -0.066$^{+0.030}_{-0.044}$ & $0.15<|Z|<0.25$ &$-$ &$-$ &$-$ &$-$ & 2012ApJ...746..149C\\
                        &                          & -0.064$^{+0.015}_{-0.004}$ & $0.25<|Z|<0.50$ &$-$ &$-$ &$-$ &$-$ & 2012ApJ...746..149C\\
                        &                          & -0.013$^{+0.009}_{-0.002}$ & $0.50<|Z|<1.00$ &$-$ &$-$ &$-$ &$-$ & 2012ApJ...746..149C\\
                        &                          & +0.028$^{+0.007}_{-0.005}$ & $1.00<|Z|<1.50$ &$-$ &$-$ &$-$ &$-$ & 2012ApJ...746..149C\\
\citet{Yong12}          & OC                       & -0.120$^{+0.010}_{-0.140}$ & $\tau \geq 2.5$; $R_{gc}<13$ &$-$ &$-$ &$-$ &$-$ & 2012AJ....144...95Y \\
\citet{Karatas12}       & MSs                      & $-$&$-$ & -$0.070\pm0.010$ &$0<z_{max}<5$ &$-$ &$-$ & 2012NewA...17...22K\\
\citet{LucKnLambert11} 	& Cepheids                 & -$0.062\pm0.002$ & $4<R_{gc}<17$ &$-$ &$-$ &$-$ &$-$ & 2011AJ....142..136L \\
\citet{Ruchti11}        & RGB,RCs,HB,MS,SG	   & +$0.010\pm0.040$ & $5<R_{gc}<10$, $|Z|<3$ &$-$ &$-$ &$-$ &$-$ & 2011ApJ...737....9R \\
\citet{Luck11}          & Cepheids	           & -$0.055\pm0.003$ & $4<R_{gc}<16$& &$-$ &$-$ &$-$ & 2011AJ....142...51L \\
\citet{Friel10}         & OC                       & -$0.076\pm0.018$ & all sample &$-$  &$-$ &$-$ &$-$ & 2010AJ....139.1942F \\
\citet{Wu09}            & OC                       & -$0.070\pm0.011$ & $6\leq R_{gc}\leq13.5$ & $-0.082\pm0.014$ &$6\leq R_{m}\leq20$ &$-$ &$-$ & 2009MNRAS.399.2146W \\
\citet{Pedicelli09}     & Cepheids                 & -$0.051\pm0.004$ & all sample &$-$ &$-$ &$-$ &$-$ & 2009A\&A...504...81P \\
\citet{Magrini09}	& OC                       & -$0.053\pm0.029$ & $7<R_{gc}<12$, $\tau \leq0.8$ Gyr&$-$ &$-$ &$-$ &$-$ &	2009A\&A...494...95M \\
\citet{Lemasle08}       & Cepheids                 & -$0.052\pm0.003$ & $5<R_{gc}<17$ &$-$ &$-$ &$-$ &$-$ & 2008A\&A...490..613L \\
\citet{Lemasle07}       & Cepheids                 & -$0.061\pm0.019$ &	$8<R_{gc}<12$ &$-$ &$-$ &$-$ &$-$ & 2007A\&A...467..283L \\
\citet{Allende06}	& FG MSs                   & 0& $1<|Z|\leq3$ & $-$ & $-$ &$-$ &$-$ &2006ApJ...636..804A \\
\citet{Nordstrom04}	& FG MSs                   & $-$&$-$ & -$0.076\pm0.014$ & $\tau\leq1.5$ &$-$ &$-$ & 2004A\&A...418..989N \\
	                &                          & $-$&$-$ & -$0.099\pm0.011$ & $4<\tau\leq6 $ &$-$ &$-$ &2004A\&A...418..989N\\
		        &                          & $-$&$-$ & +$0.028\pm0.036$ & $\tau>10 $ &$-$ &$-$ &2004A\&A...418..989N\\				
\citet{Chen03}          & OC                       & -$0.063\pm0.008$ & $6<R_{gc}<15$  & &$-$ &$-$ &$-$ & 2003AJ....125.1397C \\
\citet{Hou02}           & OC                       & -$0.099\pm0.008$ & $6.5<R_{gc}<16$ &$-$ &$-$ &$-$ &$-$ & 2002ChJAA...2...17H \\
\citet{Friel02}         & OC 	                   & -$0.060\pm0.010$ & $7<R_{gc}<16$  &$-$ &$-$ &$-$ &$-$ & 2002AJ....124.2693F \\
                        & 	                   & -$0.023\pm0.019$ & $\tau<2$  &$-$ &$-$ &$-$ &$-$ & 2002AJ....124.2693F \\
                        & 	                   & -$0.053\pm0.018$ & $2\leq \tau \leq 4$  &$-$ &$-$ &$-$ &$-$ & 2002AJ....124.2693F \\
                        & 	                   & -$0.075\pm0.019$ & $\tau>4$   &$-$ &$-$ &$-$ &$-$ & 2002AJ....124.2693F \\
\citet{Andrievsky02}    & Cepheids                 & -$0.130\pm0.030$ & inner disc &$-$ &$-$ &$-$ &$-$ & 2002A\&A...392..491A \\
\hline
\end{tabular}
\\
Abbreviations: Main Sequence Stars: MSs; Red Clump Stars: RCs; Red Giants: RG; Open Cluster: OC; Red Giant Branch: RGB; Horizontal Branch: HB; Sub-giant: SG 
}
\end{table*}
\end{landscape}

\begin{landscape}
\textwidth = 650pt
\begin{table*}
\setlength{\tabcolsep}{3pt}
\centering
{\scriptsize
\caption{Vertical metallicity gradients appeared in the literature. Distances ($d$, $R_{gc}$, $Z$, $z_{max}$) in kpc, age ($\tau$) in Gyr.}
\begin{tabular}{lcccccr}
\hline
~~~~~~~~~~~~~~Author &Tracer Object&d[Fe/H]/d$|Z|$ &Remark&d[Fe/H]/d$z_{max}$ &Remark&Bibliographic Code \\
      & &(dex~kpc$^{-1}$)& &(dex~kpc$^{-1}$)& & \\
\hline
\citet{Plevne15}         & F-G MSs          & $-$ & $-$ &-$0.176\pm0.039$ & $0<z_{max}\leq 0.825$ & 2015PASA...32...43P \\
                         &                  & $-$ & $-$ &-$0.119\pm0.036$ & $0<z_{max}\leq 1.5$   & 2015PASA...32...43P\\
\citet{Xiang15}          & Turnoff stars    & -$0.110\pm0.020$ & $\tau>11$ & $-$ & $-$ & 2015RAA....15.1209X\\
\citet{Huang15}          & RC    	    & -$0.206\pm0.006$ & $|Z|\leq1,7<R_{gc}\leq8$         &$-$ &$-$ & 2015RAA....15.1240H\\
                         &     	            & -$0.116\pm0.008$ & $|Z|\leq1,8<R_{gc}\leq9$         &$-$ &$-$ & 2015RAA....15.1240H\\
                         &     	            & -$0.052\pm0.010$ & $|Z|\leq1,9<R_{gc}\leq10$        &$-$ &$-$ & 2015RAA....15.1240H\\
                         &     	            & $~0.000\pm0.012$ & $|Z|\leq1,10<R_{gc}\leq11$       &$-$ &$-$ & 2015RAA....15.1240H\\
                         &     	            & +$0.008\pm0.008$ & $|Z|\leq1,11<R_{gc}\leq12$       &$-$ &$-$ & 2015RAA....15.1240H\\
                         &     	            & +$0.057\pm0.012$ & $|Z|\leq1,12<R_{gc}\leq13$       &$-$ &$-$ & 2015RAA....15.1240H\\
                         &     	            & +$0.047\pm0.012$ & $|Z|\leq1,13<R_{gc}\leq14$       &$-$ &$-$ & 2015RAA....15.1240H\\
\citet{Mikolaitis14}     & FGK MSs 	    & -$0.107\pm0.009$ & Thin disc (main)  &$-$ &$-$ & 2014A\&A...572A..33M\\
    			 &   	            & -$0.057\pm0.016$ & Thin disc (clean) &$-$ &$-$ & 2014A\&A...572A..33M\\
	                 &   	            & -$0.072\pm0.006$ & Thick disc (main) &$-$ &$-$ & 2014A\&A...572A..33M\\
			 &   	            & -$0.037\pm0.016$ & Thick disc (clean)&$-$ &$-$ & 2014A\&A...572A..33M\\
\citet{Schlesinger14}    & G MSs            & -$0.243^{+0.039}_{-0.053}$ &all sample& $-$&$-$ & 2014ApJ...791..112S \\
\citet{Boeche14}         & RCs              & -$0.112\pm0.007$ & $0<|Z|\leq2$ &$-$ &$-$ &2014A\&A...568A..71B \\
\citet{Hayden14}         & RG               & -$0.305\pm0.011$ & $0<|Z|\leq2$ &$-$ &$-$ & 2014AJ....147..116H \\
\citet{Bergemann14}      & FGK MSs          & -$0.068\pm0.014$ & $|Z|\leq0.3$ &$-$ &$-$ & 2014A\&A...565A..89B \\
                         &                  & -$0.114\pm0.009$ & $0.3<|Z|\leq0.8$ &$-$ &$-$   & 2014A\&A...565A..89B \\
\citet{Carrell12}        & FGK MSs          & -$0.113\pm0.010$ & $7<R_{gc}<10.5$, Isoc. method  &$-$ &$-$ &2012AJ....144..185C \\
	                 &                  & -$0.125\pm0.008$ & $7<R_{gc}<10.5$, Photo. method &$-$ &$-$ &2012AJ....144..185C \\	
\citet{Bilir12}          & RCs              & $-$ &$-$ &-$0.109\pm0.008$ & $TD/D\leq0.1$ & 2012MNRAS.421.3362B \\
                         &                  & $-$ &$-$ &-$0.260\pm0.031$ & $e_v\leq0.07$ & 2012MNRAS.421.3362B \\
                         &                  & $-$ &$-$ &-$0.167\pm0.011$ & $e_v\leq0.12$ & 2012MNRAS.421.3362B \\	
                         &                  & $-$ &$-$ &-$0.022\pm0.005$ & $e_v>0.25$ & 2012MNRAS.421.3362B \\	
\citet{Peng12}           & MSs              & -$0.210\pm0.050$ & $0<Z<2$ &$-$ &$-$ & 2012MNRAS.422.2756P \\
\citet{Katz11}           & RGB,RCs,HB,MS,SG & -$0.068\pm0.009$ & Thick disc &$-$ &$-$ & 2011A\&A...525A..90K \\
\citet{Chen11}           & RHB              & -$0.120\pm0.010$ & $0.5<|Z|<3$ &$-$ &$-$ & 2011AJ....142..184C \\
                         &                  & -$0.220\pm0.070$ & $1<|Z|<3$ &$-$ &$-$ & 2011AJ....142..184C \\
\citet{Kordopatis11}     & FGK MSs          & -$0.140\pm0.050$ & $1\leq Z\leq 4$ &$-$ &$-$ & 2011A\&A...535A.107K \\
\citet{Ruchti11}         & RGB,RCs,HB,MS,SG & -$0.090\pm0.050$ & $6<R_{gc}<10$, $|Z|<3$ &$-$ &$-$ & 2011ApJ...737....9R \\
\citet{Yaz10}            & G MSs            & -$0.320\pm0.010$ & $Z<2.5$    & $-$ &$-$ & 2010NewA...15..234Y \\		
                         &                  & -$0.300\pm0.060$ & $3<Z<5.5$  & $-$ &$-$ & 2010NewA...15..234Y \\
                         &                  & -$0.010\pm0.010$ & $6<Z<10$   & $-$ &$-$ & 2010NewA...15..234Y \\
\citet{Siegel09}         & FS               & -$0.150$         & $Z<4$ &$-$ & $-$ & 2009MNRAS.395.1569S \\
\citet{Soubiran08}       & RCs              & -$0.300\pm0.030$ & $d<1$ &$-$ & $-$ & 2008A\&A...480...91S \\
                         &                  & $-$ & $-$ &-$0.310\pm0.060$ & $0\leq z_{max}<1.2$ & 2008A\&A...480...91S \\
\citet{Ak07a}            & G MSs            & -$0.380\pm0.060$ & $3 \leq Z<5$  &$-$ &$-$ & 2007NewA...12..605A \\
                         &                  & -$0.080\pm0.070$ & $5 \leq Z<10$ &$-$ &$-$ & 2007NewA...12..605A \\	
\citet{Ak07b}            & G MSs            & -$0.160\pm0.020$ & $Z<3$ North &$-$ &$-$ & 2007AN....328..169A \\
                         &                  & -$0.220\pm0.020$ & $Z<3$ South & $-$&$-$ &2007AN....328..169A  \\
\citet{Allende06}        & FG MSs           & 0.030&$1<|Z|\leq3$; [Fe/H]$>-1.2$ &$-$  &$-$&2006ApJ...636..804A\\
\citet{Marsakov06}       & FS               & -$0.290\pm0.060$&Thin disc&$-$  &$-$& 2006A\&AT...25..157M\\
\citet{Chen03}           & OC               & $-$ &$-$ &-$0.295\pm0.050$  &$6<R_{gc}<15$ &2003AJ....125.1397C \\
\citet{Karaali03}        & G MSs            & -$0.200$ & $Z\leq8$ & $-$ & $-$ & 2003MNRAS.343.1013K \\
\citet{Barta03}          & FS               & -$0.230\pm0.040$ & $Z<1.3$ & $-$ & $-$ & 2003BaltA..12..539B \\
\hline
\end{tabular}
\\
Abbreviations: Main Sequence Stars: MSs; Red Clump Stars: RCs; Red Giants: RG; Open Cluster: OC; Red Giant Branch: RGB; Horizontal Branch: HB; Red Horizontal Branch: RHB; Sub-giant: SG; Field Stars: FS.
}
\end{table*}
\end{landscape}

\section{The Data}
The data are selected from the Radial Velocity Experiment's fourth data release \citep[DR4;][]{Kordopatis13}. RAVE DR4 is the first RAVE release that uses DENIS $I$-band magnitude for its input catalogue, instead of pseudo $I$-band magnitude.  The new catalogue presents radial velocities for 482,430 stars. The internal errors in radial velocities are reduced from $\sim$ 2 km s$^{-1}$ to 1.4 km s$^{-1}$ for 68\% of the RAVE DR4 sample in comparison with 68\% of the RAVE DR3 sample \citep{Siebert11}.  Radial velocity errors of the sample have median value at $0.80\pm0.24$ km s$^{-1}$.  One of the important features of this data release is that it uses a new pipeline with MATISSE \citep{Recio06} and DEGAS \citep{Bijaoui12} algorithms to derive more reliable stellar atmospheric parameters (effective temperature, surface gravity, and metallicity), where \citet{Kordopatis13} extends the grid used in \citet{Bijaoui12} from 4500 K down to 3000 K.  Also, the proper motions of stars combined from various source catalogues such as Tycho-2 \citep {Hog00}, UCAC2, UCAC3, UCAC4 \citep {Zacharias10,Zacharias13}, PPMX, PPMXL \citep {Roeser10} and SPM4 \citep{Girard11}, all of which covers different portions of sample stars. UCAC4 proper motions are selected as an input for kinematic parameter calculations since it covers 98\% of the RAVE DR4 data. Proper motion errors range from 0.5 to 4 mas yr$^{-1}$ and its median value is $1.69\pm0.65$ mas yr$^{-1}$. The near infra-red magnitudes are taken from the Two Micron All Sky Survey \citep[2MASS;][]{Skrutskie06}, All-Sky Catalog of Point Sources \citep[2MASS;][]{Cutri03}.  

Distances of the RAVE DR4 stars were estimated by two different procedures. One is from \citet{Zwitter10}'s method of projection of individual stellar parameters from RAVE DR4 pipelines on a set of isochrones and obtains the most likely value for a star's absolute magnitudes. The other is from \citet {Binney14}'s Bayesian distance-finding method, which is an improved version of \citet {Burnett10}'s algorithm. However, in our study the distances for our RC sample are calculated by a different procedure. In distance estimation of the RC stars we preferred near infra-red magnitudes of 2MASS. These photometric bands are not affected much from interstellar reddening and absorption. The $K_s$ absolute magnitude of the RC stars only has a weak dependence on metallicity. This relation demonstrated in many studies within the last decade \citep[see for example][and references therein]{Lopez02, Lopez04, Cabrera05, Cabrera07a, Cabrera07b, Cabrera08, Bilir12}. 

\begin{figure*}
\centering
\includegraphics[trim=1cm 2cm 2cm 1cm, clip=true, scale=0.5, angle=0]{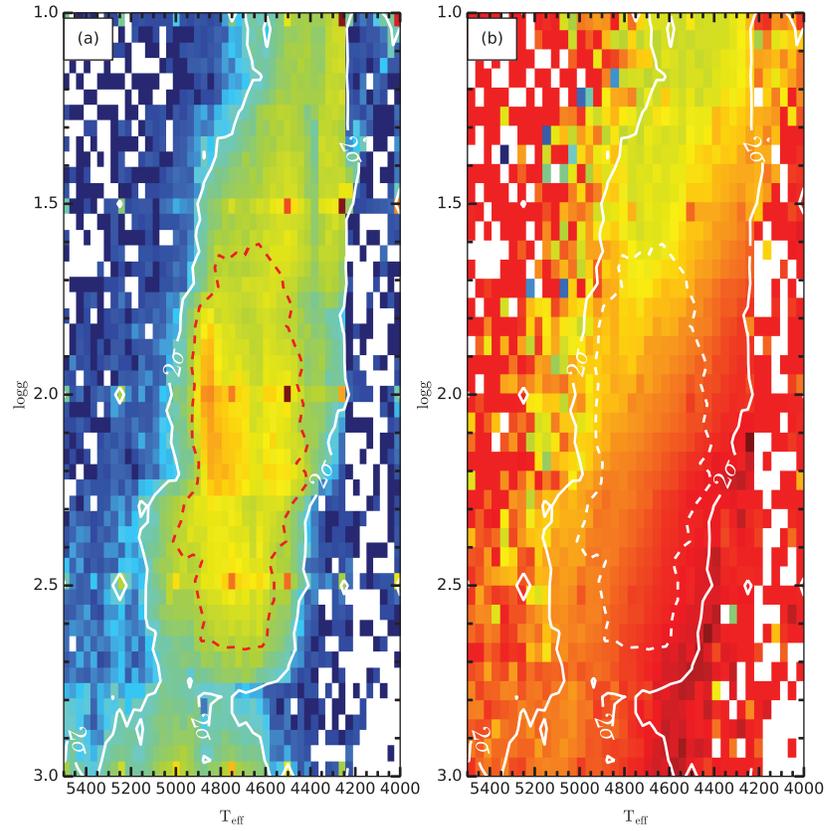}
\caption{T$_{\rm eff}-\log g$ diagram of the RC region, colour coded for logarithmic number density (a). Red dashed and white solid lines show $1\sigma$ and $2\sigma$ regions, respectively. T$_{\rm eff}-\log g$ diagram of the RC region, colour coded for metallicity (b). White dashed and solid lines show $1$ and $2\sigma$ regions, respectively.}
\label{Fig01}
\end{figure*}

In Fig. \ref{Fig01}, T$_{\rm eff}-\log g$ diagram of the region where the RC stars most likely to reside are presented as colour coded diagrams in two panels, one with logarithmic number density and the other with logarithmic metallicity. Also, 1$\sigma$ and 2$\sigma$ contour lines based on the most crowded region are drawn on the figure. Depending on the contour lines, we selected the stars that reside in the 1$\sigma$ area which correspond to 68,663 stars. The whole region covers $1<\log g<3$ and $4000<$T$_{\rm eff}(K)<5400$. Selection of suitable RC sample also depends on other constraints: (1) having a cross matched UCAC4 catalogue \citep{Zacharias13} proper motion values (since it covers the largest portion of the sample compared with other proper motion catalogues); (2) having metallicity values from the RAVE DR4 chemical pipeline \citep{Kordopatis13}; (3) according to \citet{Kordopatis13}, stars with $S/N\geq40$ in the RAVE DR4 catalogue have more accurate astrophysical parameters so this cut was made; (4) avoid repeated observations. This reduces the number of stars to 52,196.

\begin{figure*}
\centering
\includegraphics[trim=1cm .1cm .1cm .1cm, clip=true, scale=0.5, angle=0]{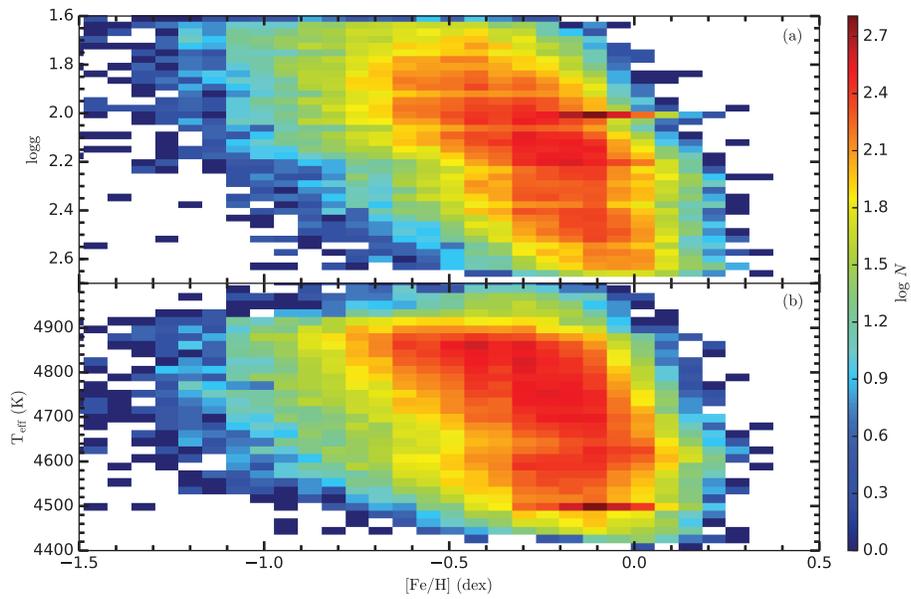}
\caption{T$_{\rm eff}$-[Fe/H] and $\log g$-[Fe/H] diagram of our RC sample.}
\label{Fig02}
\end{figure*}

Given that Fig. \ref{Fig01} is a pseudo-HR diagram, it shows that stars on the first ascent of the giant branch have brighter absolute magnitudes as $\log g$ and T$_{\rm eff}$ both decrease.  Brighter absolute magnitudes correspond to larger line-of-sight distances.  As RAVE observes away from the Galactic plane, larger line-of-sight distances means statistically larger numbers of thick disc stars and thus lower metallicities.  Fig. \ref{Fig02} shows there is a correlation between [Fe/H] and T$_{\rm eff}$ and between [Fe/H] and $\log g$ in our RC sample, which suggests it is contaminated with first ascent giants.  With a similar selection of RC stars, \citet{Williams13} showed that the first ascent giants in their RC sample have a very similar distance distribution.  Given this and that there is also a real dispersion in absolute magnitude among bona fide RC stars, such distance uncertainties will reduce our measured gradients.  Nevertheless as we looking for general trends, we consider such an effect to be second order.

\begin{figure*}
\centering
\includegraphics[trim=3cm 3cm 2cm 1cm, scale=0.45, angle=0]{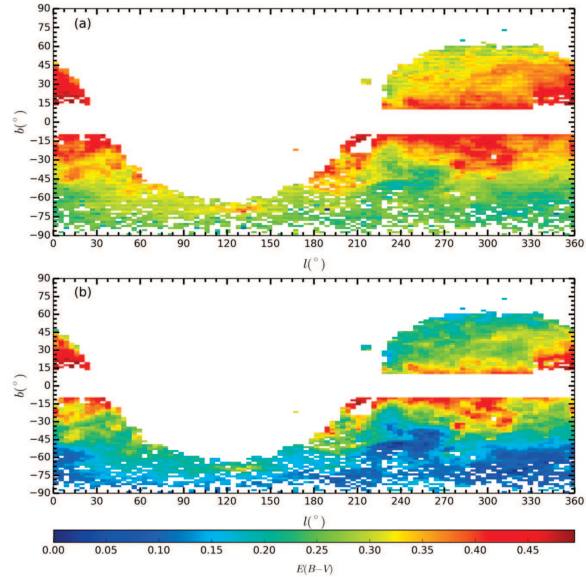}
\caption{Colour coded for $E_{\infty}(B-V)$ colour excess (a), and  reduced colour excess $E_d{(B-V)}$ (b) distributions of 52,196 stars in Galactic coordinates.} 
\label{Fig03}
\end{figure*} 

\begin{figure*}
\centering
\includegraphics[width=8cm,height=7cm, angle=0]{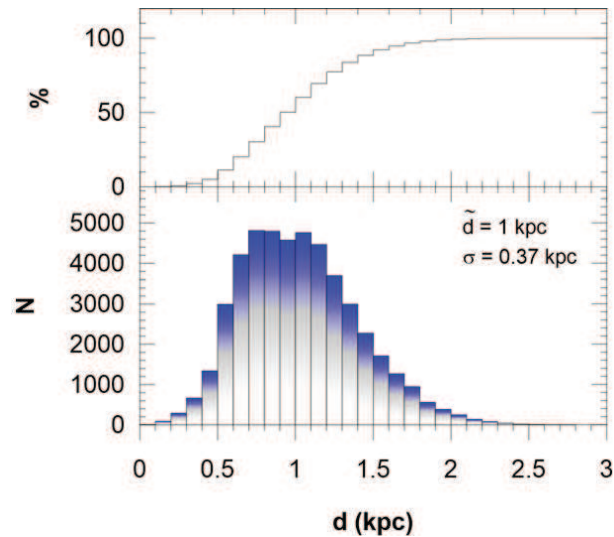}
\caption{Distance histogram of 52,196 the RC stars. Median and standard deviation of distance distribution are 1 and 0.37 kpc, respectively. The corresponding percentage of RC stars are also shown in the upper panel of the diagram.} 
\label{Fig04}
\end{figure*} 

\begin{figure*}
\centering
\includegraphics[width=6cm,height=12cm,keepaspectratio]{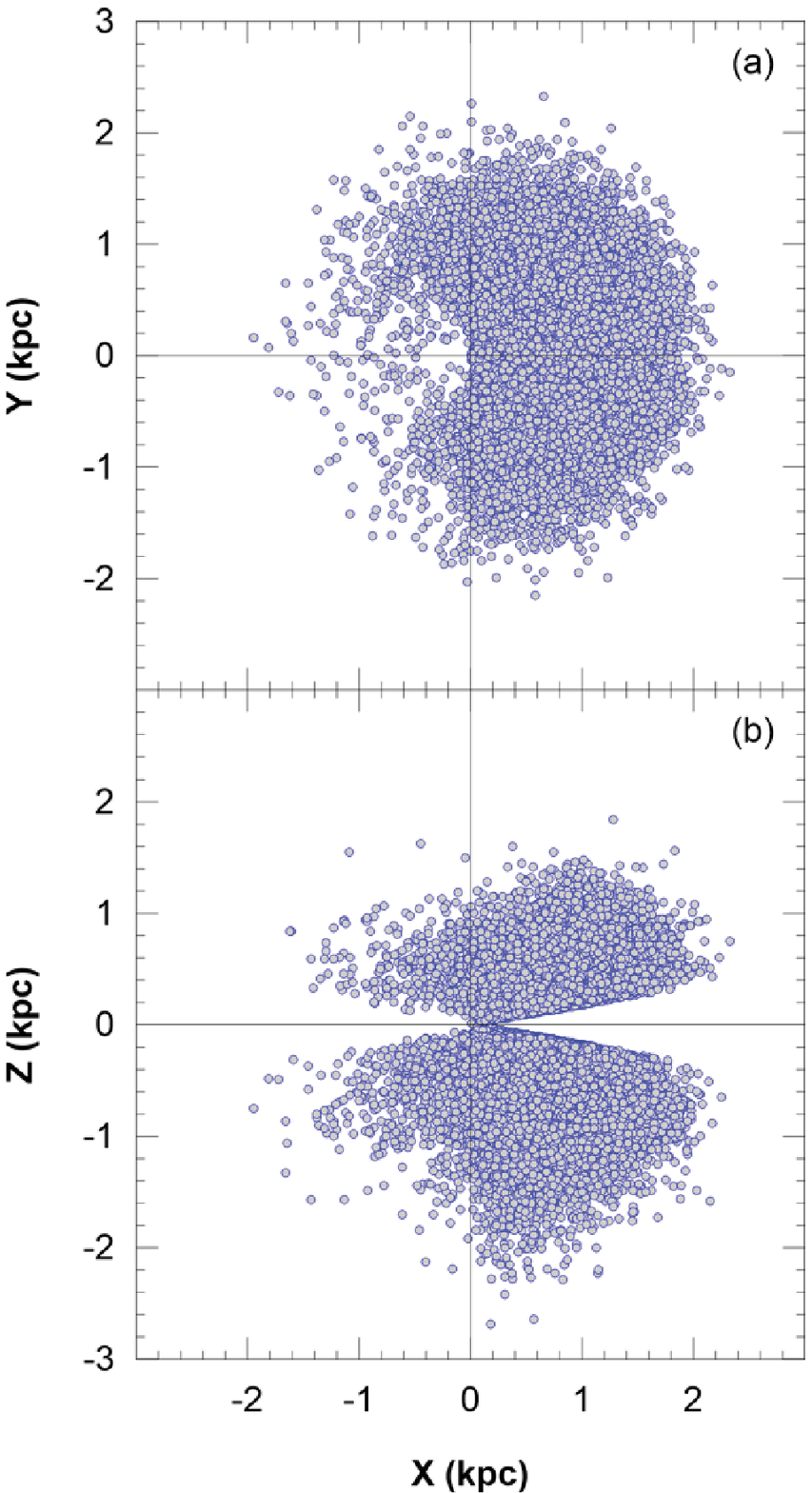}
\caption{Heliocentric distribution of the RC stars projected on $X$-$Y$ and $X$-$Z$ planes.} 
\label{Fig05}
\end {figure*}

We adopted \citet{Groenewegen08}'s $M_{K_s}=$-$1.54\pm 0.04$ mag for the sample. Using \citet{Schletal11}'s reddening maps, we evaluated the $E_{\infty}(B-V)$ colour excess for each star individually, and obtained the reduced value of colour excess $E_{d}(B-V)$ related to their respective distances of each star in the RC sample by using \citet{Bahcall80}'s equation. Further iterations were carried out for the dereddening of the $K_s$ apparent magnitude for each star \citep[cf.][]{Cokkun11,Cokkun12}. In Fig. \ref{Fig03}, panels (a) and (b) represent the original colour excess ($E_{\infty}(B-V)$) and the reduced colour excess $E_{d}(B-V)$ of the RC stars as colour coded in Galactic coordinates, respectively. 

The distance histogram of the sample stars is presented in Fig. \ref{Fig04}. The median distance and it's standard deviation are $d=1$ and $\sigma=0.37$ kpc, respectively.  The median value for the relative distance errors is 0.05 kpc. Heliocentric space distributions of the RC stars are shown in Fig. \ref{Fig05} in two panels. Most of the stars are found in the first and fourth Galactic quadrants. The median distance values of the sample in $X$, $Y$ and $Z$ are 0.67, 0.09, and -0.24 kpc, respectively. Our sample covers a distance range of $0<|Z|<3$ kpc and $5.5<R_{gc}<11$ kpc. However, most of our sample stars (83.6\%) are concentrated in $7<R_{g}<9$ kpc.

RAVE DR4 line-of-sight velocities \citep{Kordopatis13}, UCAC4 proper motions \citep{Zacharias13} and photometric distances are combined to obtain space velocity components ($U$, $V$, $W$) of 52,196 RC stars, which are calculated with \citet{Johnson87}'s standard algorithms and the transformation matrices of a right handed system for epoch J2000 (as described in the International Celestial Reference System of the {\it Hipparcos} and Tycho-2 Catalogues \citep{ESA97}. Hence, $U$, $V$ and $W$ are the components of a velocity vector of a star with respect to the Sun, where $U$ is positive towards the Galactic centre ($l=0^{o}$, $b=0^{o}$), $V$ is positive in the direction of Galactic rotation ($l=90^{o}$, $b=0^{o}$) and $W$ is positive towards the North Galactic Pole ($b=90^{o}$). The Galactic rotational velocity of the Sun is adopted as 222.5 km s$^{-1}$ \citep{Schonrich12}. Since stars in our Galaxy orbit around the Galactic center with different speeds, \citet{Mihalas81} suggested a series of corrections on $U$ and $V$ space velocities in order to compensate for this. The $W$ space velocity is not affected by this behavior thus needed no correction. The first order Galactic differential rotation corrections are -$61.29<dU<34.12$ and -$3.76<dV<5.34$ km s$^{-1}$. Then, space velocities are reduced by applying solar local standard of rest values (LSR) of \citet{Cokkun11} for all stars $(U_{\odot}, V_{\odot}, W_{\odot})_{LSR}=(8.83\pm 0.24, 14.19\pm 0.34, 6.57\pm 0.21)$ km s$^{-1}$ to LSR.

\begin{figure}
\centering
\includegraphics[width=8 cm,height=16cm,keepaspectratio]{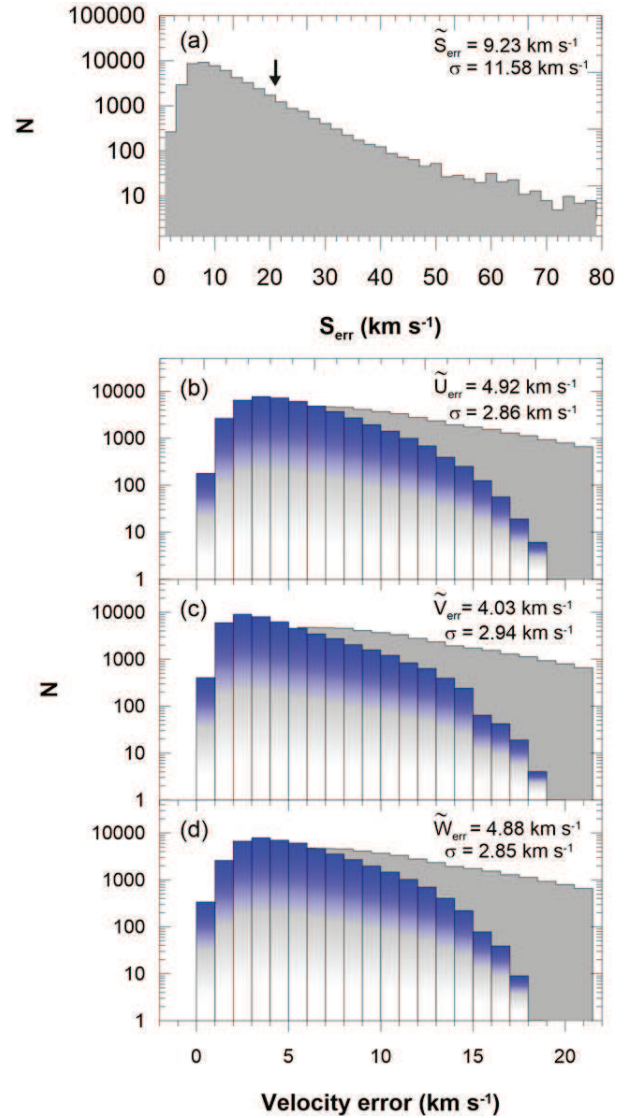}
\caption{Total space velocity error histogram of 52,196 stars ({\it top panel}). The median and standard deviation are 9.23 km s$^{-1}$ and 11.58 km s$^{-1}$, respectively, which of their sum gives approximately 21 km s$^{-1}$ and this is the last constraint that applied to the sample. As a result 47,406 RC stars are remained as the final sample. Histograms of $U$, $V$, and $W$ space velocity errors in comparison with total space velocity error histogram of 47,406 RC stars ({\it lower panels}).} 
\label{Fig06}
\end {figure}

\begin{figure}
\centering
\includegraphics[width=8cm,height=16cm,keepaspectratio]{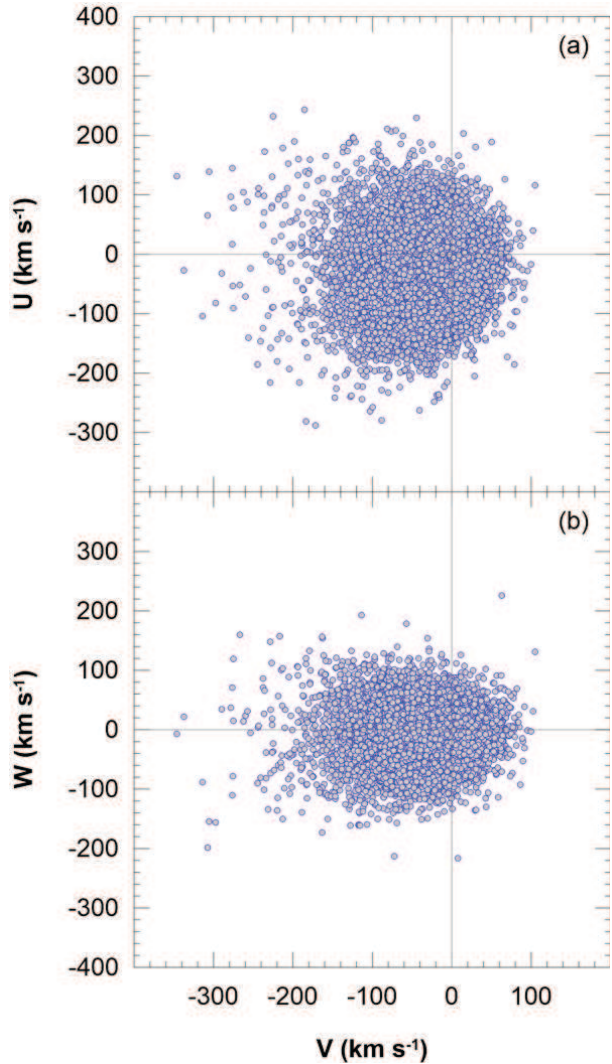}
\caption{Distribution of space velocity components of the RC stars in $V$-$U$ and $V$-$W$ planes.} 
\label{Fig07}
\end {figure}

\begin{figure}
\centering
\includegraphics[width=8cm,height=4cm,keepaspectratio]{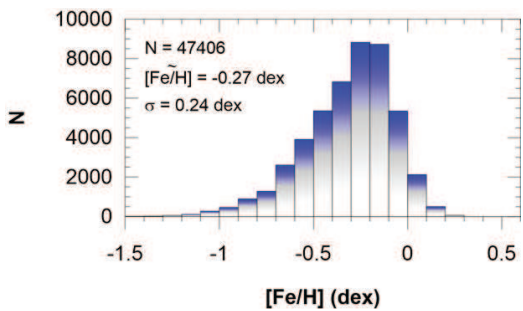}
\caption{Metallicity histogram of 47,406 RC stars. Median metallicity of the distribution is [Fe/H]=-0.27 dex and its standard deviation $\sigma_{[Fe/H]}=0.24$ dex.} 
\label{Fig08}
\end {figure}

Using \citet{Johnson87}'s algorithm, the uncertainties of space velocities are calculated by propagating the uncertainties in radial velocity, proper motion and distance. Kinematic input parameter errors vary in $0.3< \gamma<6.6$ km s$^{-1}$, $0.5<\mu<7$ mas yr$^{-1}$, and $0.01<\sigma_{\pi}/\pi <0.13$ intervals for radial velocity, proper motion and distance, respectively and their corresponding medians and standard deviations are 0.80$\pm$0.24, 1.69$\pm$0.65 and 0.05$\pm$0.26, respectively. The total space velocity errors are calculated as a square root of individual space velocity errors of the RC stars, i.e. ($S_{err}^2=U_{err}^2+V_{err}^2+W_{err}^2$). The distribution of total space velocities of 52,196 RC stars is presented in Fig. \ref{Fig06}a. We applied a final constraint on total space velocity errors of the RC stars. In order to remove the most discrepant stars, we applied a cut-off point of 21 km s$^{-1}$ which is deduced from the stars within 1-$\sigma$ prediction in total space velocity error. Thus, our final sample is reduced to 47,406 RC stars. The final sample has the median space velocity errors and standard deviations of ($U_{err}$, $V_{err}$, $W_{err}$)=(-$4.92\pm2.86$, $4.03\pm 2.94$, $4.88\pm 2.85$) km s$^{-1}$. In Fig. \ref{Fig06}b-d the histograms of space velocity errors for the final RC sample are shown. Figs. \ref{Fig07} and \ref{Fig08} show the space velocity and metallicity distributions of the final RC sample respectively. 
Using \citet{Johnson87}'s algorithm, the uncertainties of space velocities are calculated by propagating the uncertainties in radial velocity, proper motion and distance. Kinematic input parameter errors vary in $0.3< \gamma<6.6$ km s$^{-1}$, $0.5<\mu<7$ mas yr$^{-1}$, and $0.01<\sigma_{\pi}/\pi <0.13$ intervals for radial velocity, proper motion and distance, respectively and their corresponding medians and standard deviations are 0.80$\pm$0.24, 1.69$\pm$0.65 and 0.05$\pm$0.26, respectively. The total space velocity errors are calculated as a square root of individual space velocity errors of the RC stars, i.e. ($S_{err}^2=U_{err}^2+V_{err}^2+W_{err}^2$). The distribution of total space velocities of 52,196 RC stars is presented in Fig. \ref{Fig06}a. We applied a final constraint on total space velocity errors of the RC stars. In order to remove the most discrepant stars, we applied a cut-off point of 21 km s$^{-1}$ which is deduced from the stars within 1-$\sigma$ prediction in total space velocity error. Thus, our final sample is reduced to 47,406 RC stars. The final sample has the median space velocity errors and standard deviations of ($U_{err}$, $V_{err}$, $W_{err}$)=(-$4.92\pm2.86$, $4.03\pm 2.94$, $4.88\pm 2.85$) km s$^{-1}$. In Fig. \ref{Fig06}b-d the histograms of space velocity errors for the final RC sample are shown. Figs. \ref{Fig07} and \ref{Fig08} show the space velocity and metallicity distributions of the final RC sample respectively. 

\begin{figure}
\centering
\includegraphics[width=8cm,height=12cm]{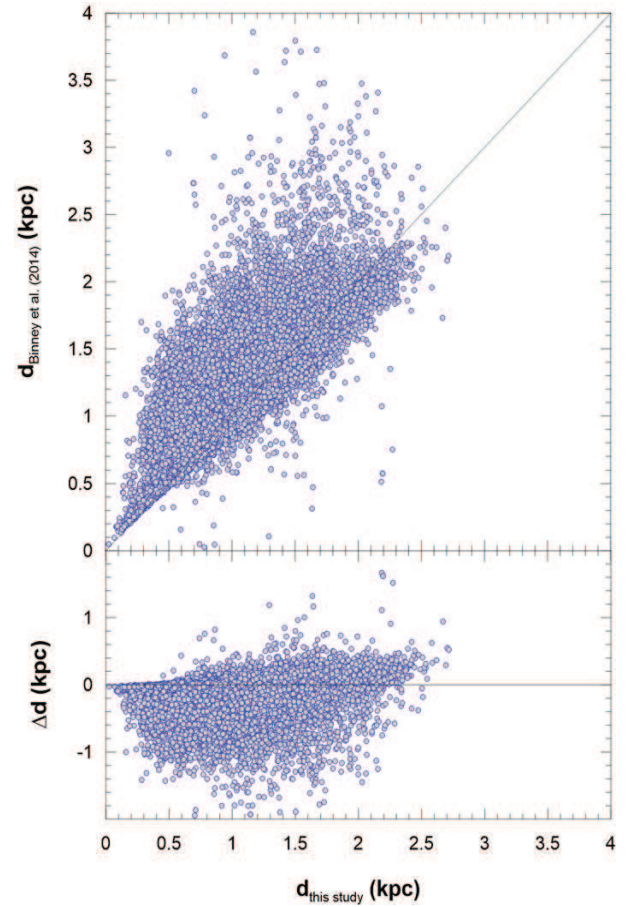}
\caption{Distance comparison of this study and \citet{Binney14}.} 
\label{Fig09}
\end {figure}

\citet{Binney14} find that their absolute magnitudes peak at $M_{K_s}$=-1.53 mag (see their Fig. 10). As we have used an almost identical value with them ($M_{K_s}$=-1.54$\pm$ 0.04 mag), we expect our distances to be very similar to the \citet{Binney14} distances. Fig. \ref{Fig09} compares the distances of our final RC sample to the \citet{Binney14}'s in RAVE DR4. While Fig. \ref{Fig09} shows that the \citet{Binney14} distances are systematically larger, they are still, as expected very similar: the mean and standard deviation of the distance difference between the two methods is 87 pc and 220 pc respectively. Out of our 47,406 RC stars, RAVE DR4-\citet{Binney14} provides distances to 41,221 i.e. it is missing 13\% of the sample. Given that our distances are very similar to the \citet{Binney14} distances and provide a larger sample, we have used our photometric parallax estimation for our metallicity gradient analysis.

Dynamical properties of individual stars in the RC sample are calculated via {\it MWPotential2014} model in {\em galpy} Galactic dynamics library of \cite{Bovy15}, which is given in detail in Table 1 of his paper. Test-particle integration for the combined Milky Way potential (for bulge, disc and halo) over 3 Gyrs is performed to obtain the Galactic orbital parameters of the closed orbit for a given star. Celestial equatorial coordinates, distances, proper motions and radial velocities of our stars are used as input parameters in the calculations of the Galactic orbital parameters such as apogalactic and perigalactic distances, maximum vertical distance from the Galactic plane, orbital angular momenta and eccentricities are obtained as output parameters.

Gradient analysis are performed both for the current orbital status of the stars and for their complete orbits. Thus, various radial and vertical distances are considered in calculations such as current Galactocentric distance ($R_{gc}$), current absolute vertical distance from the Galactic plane ($|Z|$), mean Galactocentric distance ($R_m$), which is the arithmetic mean of the perigalactic ($r_p$) and apogalactic ($r_a$) distances of a star's orbit, and maximum vertical distances from the Galactic plane towards the North ($z_{max}$) and South Galactic ($z_{min}$) poles of a star's entire orbit. Moreover, general distribution of stars on the planar and vertical eccentricity planes are investigated. Planar eccentricity is an output from the {\em galpy} dynamical library ($e_p=\frac{r_a-r_p}{r_a+r_p}$), while vertical eccentricities are calculated as $e_v=\frac{\frac{1}{2}(|z_{max}|+|z_{min}|)}{R_m}$.

\section{Results}
The metallicity gradients of our RC stars are considered in two general regimes: one for their current orbital positions and one from orbital properties calculated for 3 Gyrs with 2 Myr steps, which give stars to have enough time to close their orbits. In order to calculate metallicity gradients, linear fits are applied to the data for each selected sub-sample. 

\subsection{Metallicity Gradients from the Current Positions of the Stars}
Even though RAVE is a shallow sky survey ($9<I<13$), RC stars have such bright absolute magnitudes ($M_{Ks}$=-1.54 mag) that these stars reach up to 3 kpc distance from the Sun in our sample (see Fig. 3). Thus, calculations of radial and vertical metallicity gradients are applicable from their current positions in the Galaxy. 

\begin{figure}
\centering
\includegraphics[width=8cm,height=10.5cm]{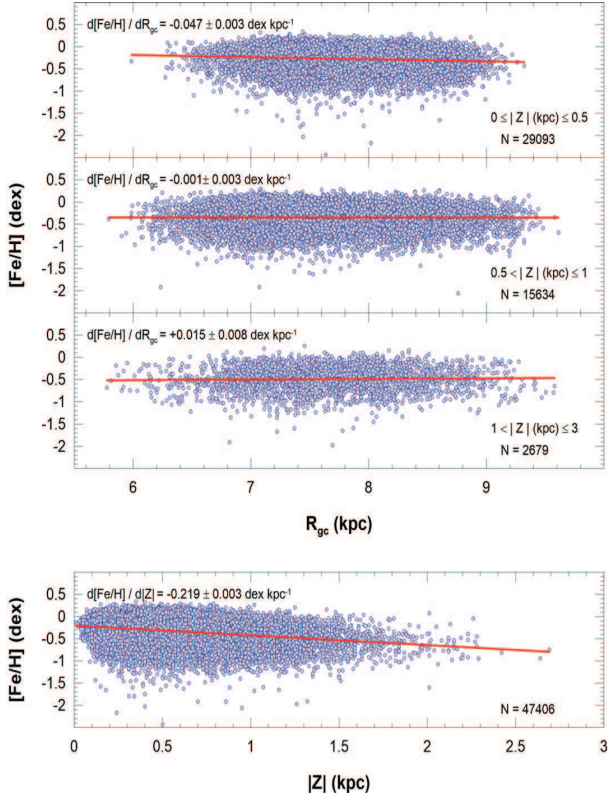}
\caption{Radial metallicity gradients for current Galactocentric distance of 47,406 RC stars in three $|Z|$ intervals, i.e. $0\leq |Z|\leq 0.5$, $0.5<|Z|\leq1$, and $1<|Z|\leq3$ kpc ({\it upper panels}). Vertical metallicity gradients for current distance from the Galactic plane of the same sample ({\it lower panel}).} 
\label{Fig10}
\end {figure}

In order to calculate radial metallicity gradients of the final RC star sample, the relation between $R_{gc}$ and [Fe/H] are analyzed for three $|Z|$ intervals, i.e. $0\leq|Z|\leq0.5$, $0.5<|Z|\leq1$, and $1<|Z|<3$ kpc. The resulting gradients are listed in Table 3 and shown in Fig. \ref{Fig10}. Radial metallicity gradients vary from -0.047$\pm$0.003 to +0.015$\pm$ 0.008 dex kpc$^{-1}$ with increasing $|Z|$ intervals. A similar trend is also seen in median metallicities of the sub-samples, which also vary from -0.26 to -0.42 dex in Table 3 (column 3). These results show that the radial gradients become flatter as the $|Z|$ distances become larger which is an expected result because as $|Z|$ increases more and more stars are members of the thick disc, which presents no radial metallicity gradient.

Analogously, vertical metallicity gradients that are obtained from the relation between $|Z|$ and [Fe/H] of the RC star sample is presented in Fig. \ref{Fig10} (lower panel), and it's resulting value is -0.219$\pm$0.003 dex kpc$^{-1}$, which shows that there is a significant vertical metallicity gradient within the Galactic disc in $0<|Z|<3$ kpc distance interval.

\begin{table}[!htb]
\centering
\caption{Radial metallicity gradients of the RC stars in each $Z$ interval for current orbital positions.}  
\begin{tabular}{cccc}
\hline
$|Z|$ interval& $N$ &  $\langle$[Fe/H]$\rangle$ &    d[Fe/H]/d$R_{gc}$\\
         (kpc)&     &    (dex)    &   (dex~kpc$^{-1}$)\\
\hline
  0.0-0.5 & 29093 & -0.26 & -0.047$\pm$0.003\\
  0.5-1.0 & 15635 & -0.29 & -0.003$\pm$0.001\\
  1.0-3.0 & ~2678 & -0.42 & +0.015$\pm$0.008\\
\hline
\end{tabular}
\end{table}

\subsection{Metallicity Gradients from the Galactic Orbits of Stars}
\begin{figure}
\centering
\includegraphics[width=8cm,height=10.67cm,keepaspectratio]{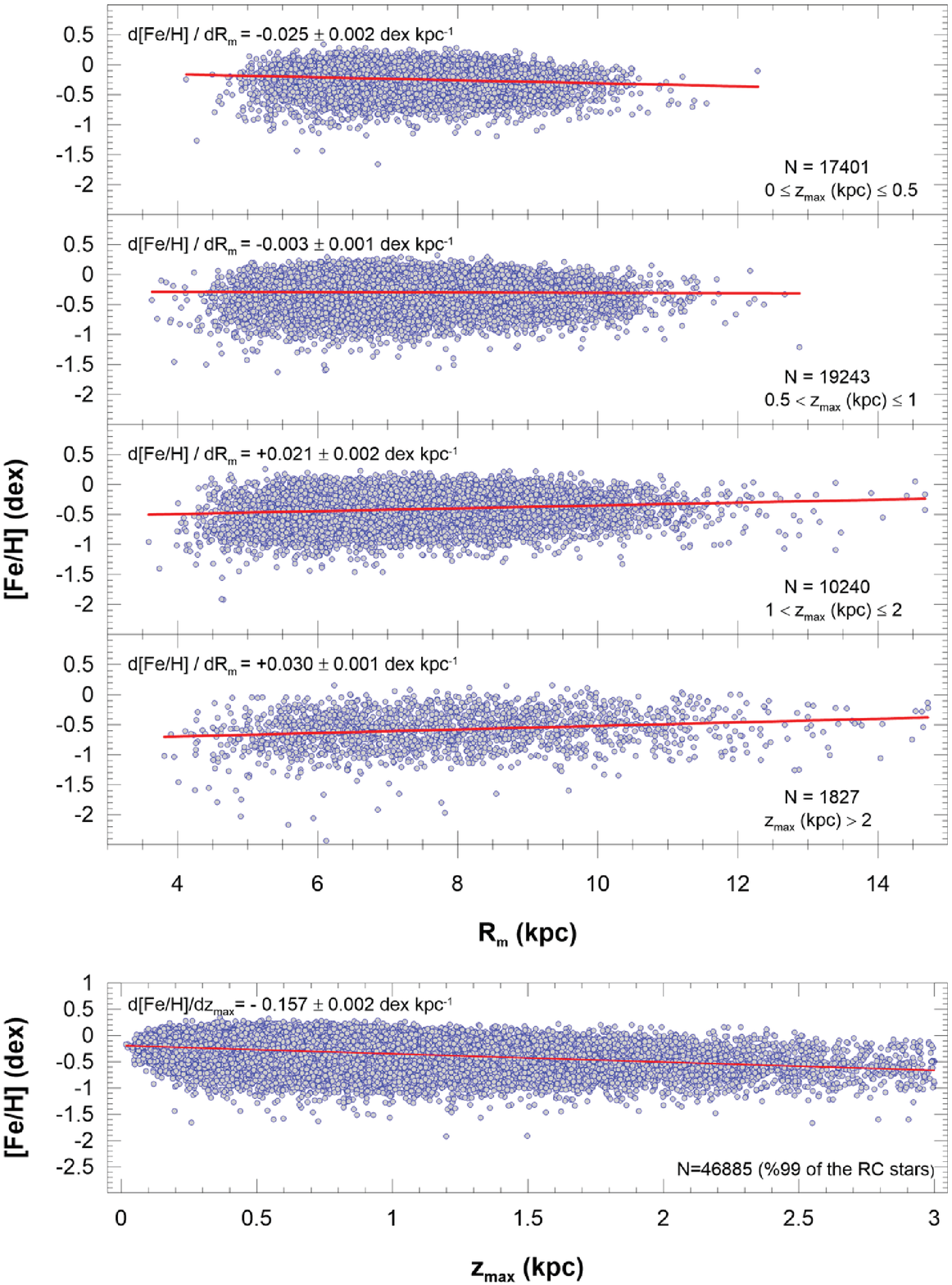}
\caption{Radial metallicity gradients for mean Galactocentric distances of 47,406 RC stars in four $z_{max}$ intervals, i.e. $0\leq z_{max}\leq0.5$, $0.5<z_{max}\leq1$, $1<z_{max}\leq2$, and $z_{max}>2$ kpc ({\it upper panels}). Vertical metallicity gradient for maximum distance from the Galactic plane of the same sample ({\it lower panel}).} 
\label{Fig11}
\end {figure}

Galactic orbital parameters of the RC stars are used in order to obtain $R_m$ and $z_{max}$ of a given star. Then, radial and vertical gradients are calculated as functions of these parameters accordingly. The resulting values of the radial metallicity gradient are listed in Table 4 and shown in Fig. \ref{Fig11} for four $z_{max}$ intervals, i.e. $0\leq z_{max}\leq0.5$, $0.5<z_{max}\leq1$, $1<z_{max}\leq2$, and $z_{max}>2$ kpc. Overall radial gradients change between -0.025$\pm$0.002 and +0.030$\pm$0.004 dex kpc$^{-1}$. Similarly median metallicities of the sub-samples vary from -0.22 to -0.57 dex with increasing $z_{max}$ in Table 4 (column 3). Sign of the gradients change from negative to positive for $z_{max}>1$ kpc which is generally considered as the domain of thick-disc stars where no or positive gradients are expected. According to \citet{Juric08}, $Z=1$ kpc is the transition region where the thin-disc component loses density in the number of stars and the thick-disc population becomes dominant.

\begin{table}[!htb]
\centering
\caption{Radial metallicity gradients of the RC stars in each $z_{max}$ interval for complete stellar orbits.}
\begin{tabular}{cccc}
\hline
$z_{max}$ interval & $N$ & $\langle$[Fe/H]$\rangle$ & d[Fe/H]/d$R_m$  \\
              (kpc)&     & (dex)      & (dex~kpc$^{-1}$)\\
\hline
     0-0.5 & 17401 & -0.22 & -0.025$\pm$0.002\\
     0.5-1 & 19243 & -0.27 & -0.003$\pm$0.001\\
     1-2   & ~8935 & -0.39 & +0.024$\pm$0.002\\
     $>$2  & ~1827 & -0.57 & +0.030$\pm$0.004\\
\hline
\end{tabular}
\end{table}

Although the $z_{max}$ distances of the RC stars can be as large as 16 kpc in the sample, 99\% of them lie within $z_{max}\leq 3$ kpc interval. Hence, we evaluated the vertical metallicity gradients for this range. Our result, -0.157$\pm$0.002 dex kpc$^{-1}$, indicates a steep metallicity gradient in the vertical direction for the Galactic disc. However, the metallicity gradient that is obtained for $|Z|$ distances in Section 3.1 is steeper than the one obtained for $z_{max}$, i.e. -0.219$\pm$0.003 dex kpc$^{-1}$.

\subsection{Metallicity Gradients According to Eccentricities and Their Implications}
Instead of assigning populations to the RC stars with any known methods, i.e. dynamical, chemical or statistical, we considered the Galactic disc as a whole and see how the stellar properties such as metallicity, space velocities, eccentricity, orbital angular momenta etc., vary with distance. Variations in metallicity gradients are examined with an additional constraint, i.e. the orbital eccentricities, to the aforementioned $z_{max}$ intervals.

\begin{figure*}
\centering
\includegraphics[width=16cm,height=10cm]{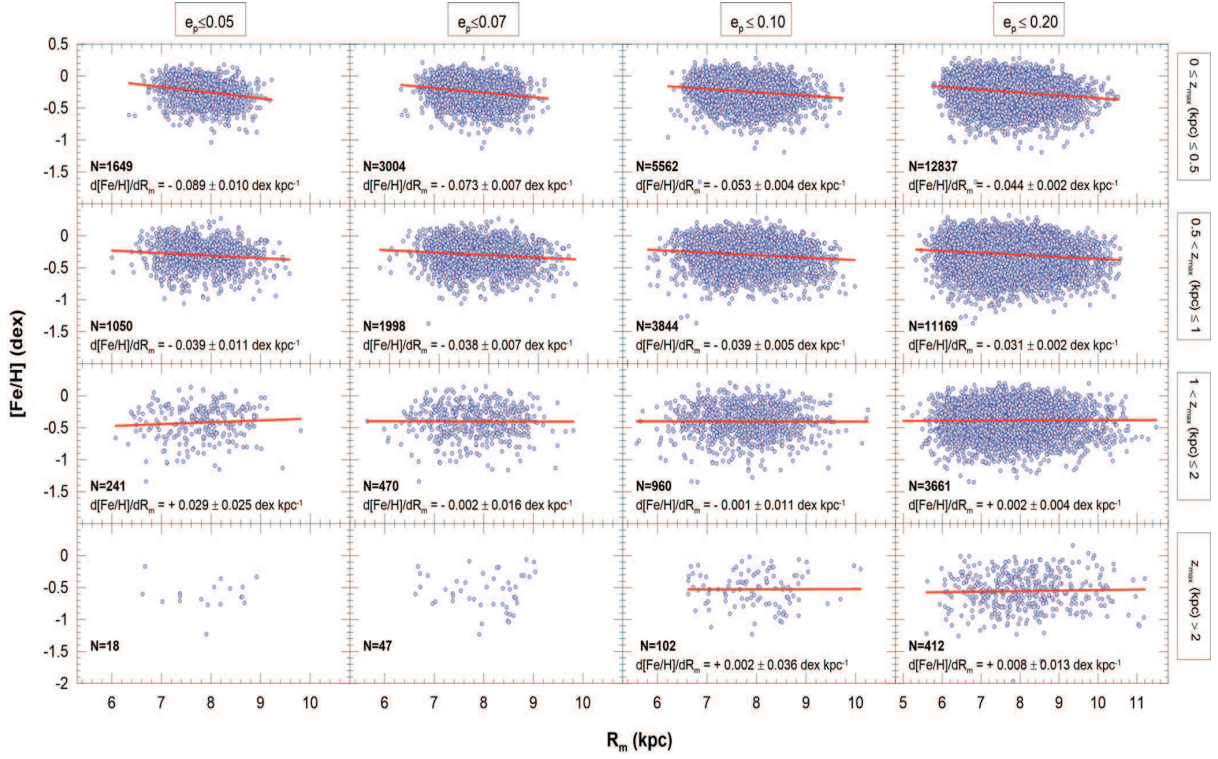}
\caption{Radial metallicity gradients for $R_m$ of 47,406 RC stars for consecutive $e_p$ sub-samples, i.e. $e_p\leq0.05$, $e_p\leq0.07$, $e_p\leq0.10$ and  $e_p\leq0.20$, in four $z_{max}$ intervals. Metallicity gradients and number of stars in each corresponding interval are also shown in the panels.} 
\label{Fig12}
\end {figure*} 

\begin{table*}[!htb]
\setlength{\tabcolsep}{4pt}
\centering
\caption{Radial metallicity gradients of the RC stars for consecutive $e_p$ limits, i.e. $\leq$0.05, 0.07, 0.10 and 0.20, in each $z_{max}$ intervals.}
\begin{tabular}{ccccccc}
\hline
 & & & $e_p\leq 0.05$ & & &$e_p\leq 0.07$\\
\hline
$z_{max}$ interval & $N$ & $\langle$[Fe/H]$\rangle$ & d[Fe/H]/d$R_m$ & $N$ & $\langle$[Fe/H]$\rangle$ & d[Fe/H]/d$R_m$\\
      (kpc)&  &(dex)&(dex~kpc$^{-1}$) &   &(dex)&(dex~kpc$^{-1}$)\\
\hline
0-0.5 & 1649 & -0.23 & -0.089$\pm$0.010 & 3004 & -0.23 & -0.073$\pm$0.007\\
0.5-1 & 1050 & -0.28 & -0.039$\pm$0.011 & 1998 & -0.27 & -0.038$\pm$0.007\\
  1-2 & ~~241& -0.39 & +0.029$\pm$0.025 & ~470 & -0.39 & -0.002$\pm$0.016\\
$>2$  & ~~~18 & -- & -- & ~~~47   & -- & -- \\     
\hline
 & & &$e_p\leq 0.10$&  &  &$e_p\leq 0.20$\\
\hline
$z_{max}$ interval & $N$ & $\langle$[Fe/H]$\rangle$ & d[Fe/H]/d$R_m$ & $N$ & $\langle$[Fe/H]$\rangle$ & d[Fe/H]/d$R_m$\\
      (kpc)&  &(dex)&(dex~kpc$^{-1}$) &   &(dex)&(dex~kpc$^{-1}$)\\
\hline
0-0.5 & 5562 & -0.23 & -0.053$\pm$0.004 & 12837 & -0.22 & -0.044$\pm$0.002\\
0.5-1 & 3844 & -0.27 & -0.039$\pm$0.005 & 11169 & -0.26 & -0.031$\pm$0.002\\
1-2   & ~960 & -0.40 & -0.001$\pm$0.011 & ~3661 & -0.37 & +0.002$\pm$0.004\\
$>2$  & ~102 & -0.52 & +0.020$\pm$0.036 & ~~412 & -0.55 & +0.008$\pm$0.013\\      
\hline
\end{tabular}
\end{table*}

\subsubsection{Radial Metallicity Gradients}
Table 5 and Fig. \ref{Fig12} give the radial metallicity gradients that we found for four $z_{max}$ intervals. As $e_p$ values increase from 0.05 to 0.20 in the $0 \leq z_{max} \leq0.5$ kpc distance interval, the radial metallicity gradients vary between -0.089$\pm$ 0.010 and -0.044$\pm$ 0.002 dex kpc$^{-1}$, respectively. Unlike the first $z_{max}$ interval, radial metallicity gradients tend to remain around the same $e_p$ limitation for $0.5<z_{max}\leq1$ kpc interval. In the remaining $z_{max}$ distance intervals, $1<z_{max}\leq2$ and $z_{max}>2$ kpc, the radial gradients show no trends. However, when all the RC stars in $0\leq z_{max} \leq0.5$ kpc interval are considered, the radial gradient slope becomes -0.025$\pm$0.002 dex kpc$^{-1}$.

\begin{figure}
\centering
\includegraphics[width=7cm,height=21cm, keepaspectratio]{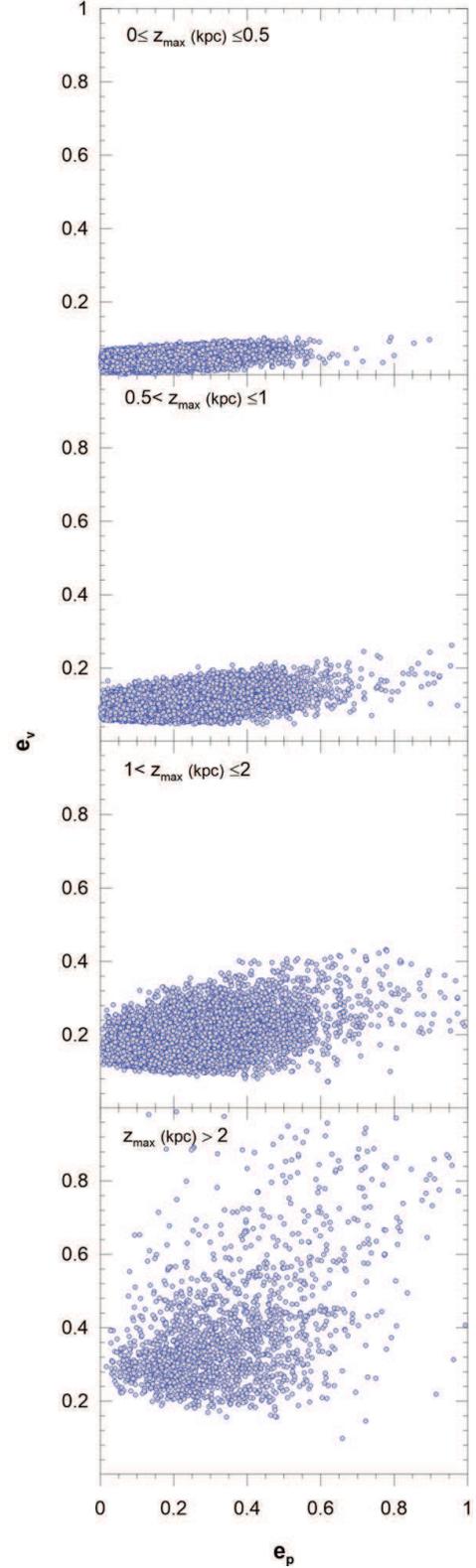}
\caption{$e_p$ - $e_v$ diagram of the RC stars for four $z_{max}$ intervals.} 
\label{Fig13}
\end {figure} 

In Fig. \ref{Fig13}, planar versus vertical eccentricity distribution of 47,406 RC stars are plotted in four $z_{max}$ intervals, i.e. $0\leq z_{max}\leq0.5$, $0.5<z_{max}\leq1$, $1<z_{max}\leq2$, and $z_{max}>2$ kpc. The median value of eccentricities change from 0.14 to 0.43 in $e_p$ as higher $z_{max}$ intervals are considered.  Young thin disc stars are predominately on circular orbits ($e_p$ close to 0) with $z_{max}$ close to the Galactic plane. Older thin disc stars have slightly higher $e_p$ and $z_{max}$.  Thick disc stars in RAVE have an asymmetric peak at $e_p=0.2$ with a relatively smooth falloff towards higher eccentricities \citep{wilson11}. Therefore the closest $z_{max}$ interval to the Galactic plane is dominated by thin disc stars.  As the $z_{max}$ interval increases, the ratio of the thin disc stars to thick disc stars decreases, so $e_p$ also increases. The median value of eccentricities change from 0.09 to 0.68 in $e_v$ as higher $z_{max}$ intervals are considered. This is because $e_v$ is proportional to $z_{max}$ by definition.  

In Tables 3 and 4, the radial metallicity gradient in the interval of 0.5-1 kpc interval from the plane is essentially flat. However, when this same interval is limited to low $e_p$, there is a negative gradient that remains the same when including more and more higher $e_p$ stars. This suggests that stars with $e_p>0.2$ have positive gradients so they cancel out the negative gradients of stars with $e_p<0.2$.

\subsubsection{Vertical Metallicity Gradients}
\begin{figure}
\centering
\includegraphics[width=8cm,height=8cm]{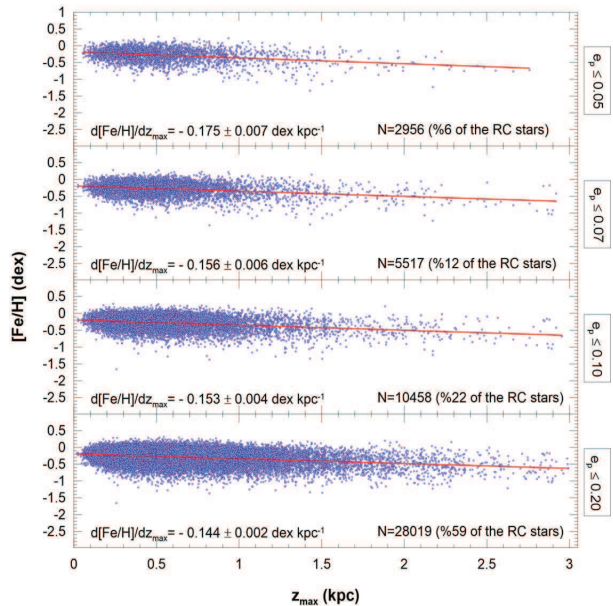}
\caption{Vertical metallicity gradients for $z_{max}$ of 47,406 RC stars for consecutive $e_p$ sub-samples, i.e. $e_p\leq0.05$, $e_p\leq0.07$, $e_p\leq0.10$ and $e_p\leq0.20$. Metallicity gradients and number of stars in each corresponding interval are also shown in the panels.} 
\label{Fig14}
\end {figure}

Vertical metallicity gradients of RC stars for four $e_p$ sub-samples are calculated regardless of $z_{max}$ intervals and shown in Table 6 and Fig. \ref{Fig14}. Gradient values change between -0.175$\pm$ 0.007 and -0.144$\pm$0.002 dex kpc$^{-1}$ and so get shallower as $e_p$ increases from 0.05 to 0.20. \citet{Mikolaitis14} found that their vertical metallicity gradient becames shallower going from their thin disc sample to their thick disc sample. Given that thick disc stars are generally more eccentric than thin disc stars, our $e_p$ cut progressively lets in more thick disc stars and so is consistent with \citet{Mikolaitis14}.

\begin{table}[!htb]
\setlength{\tabcolsep}{5pt}
\centering
\caption{Vertical metallicity gradients of the RC stars for consecutive $e_p$ limits, i.e. $\leq$0.05, 0.07, 0.10 and 0.20, in $0<z_{max}\leq3$ kpc interval.}
\begin{tabular}{rcc}
\hline
$e_p$ range   & $N$ & d[Fe/H]/d$z_{max}$\\
              &     &   (dex~kpc$^{-1}$)\\
\hline
$e_p\leq0.05$ &  2956 & -0.175$\pm$0.007 \\
$e_p\leq0.07$ &  5517 & -0.156$\pm$0.006 \\
$e_p\leq0.10$ & 10458 & -0.153$\pm$0.004 \\
$e_p\leq0.20$ & 28019 & -0.144$\pm$0.002 \\
\hline
\end{tabular}
\end{table}

\subsection{Comparing Radial Metallicity Gradients using $R_{gc}$ and $R_m$}
\begin{figure}
\centering
\includegraphics[width=8cm,height=8cm]{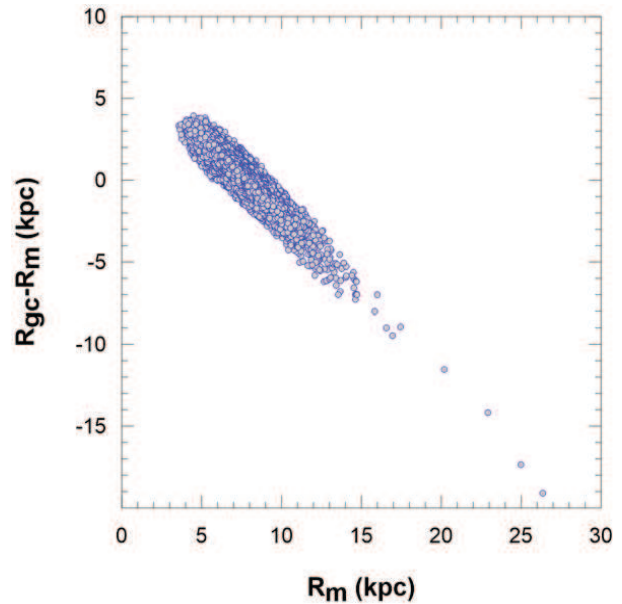}
\caption{$R_m$ and $R_{gc}$-$R_m$ diagram.} 
\label{Fig15}
\end {figure}

Tables 3 and 4 show that the radial metallicity gradient of the RC stars is steeper when considering the current position of the stars ($R_{gc}$) than that when considering their mean orbit ($R_m$) in the closest $z_{max}$ to the Galactic plane. One possible interpretation of this is due to a selection effect at the boundaries of our observed regions.  At the inner Galactic boundaries of our $R_{gc}$ sample, stars with $R_{gc}$-$R_m >0$ can be included in our sample (see Fig. \ref{Fig15}). These are higher $e_p$ stars from the inner region that are more metal rich than our $R_{gc}$-defined sample. Inclusion of these stars in our sample increases the [Fe/H] at the inner Galactic boundaries of our $R_{gc}$ sample.

Similarly, at the outer Galactic boundaries of our $R_{gc}$ sample, stars with $R_{gc}$-$R_m<0$ can be included in our sample (see Fig. \ref{Fig15}). These are also higher $e_p$ stars from the outer region that are more metal poor than our $R_{gc}$-defined sample. Inclusion of these stars in our sample decreases the [Fe/H] at the outer Galactic boundaries of our $R_{gc}$ sample. These boundary effects steepen the [Fe/H] gradient of our $R_{gc}$-defined sample.  

Our $R_m$-defined sample extends over a wider range of Galactic radii (cf. Figs. 8 and 9) and by averaging over $r_a$ and $r_p$ by definition, this sample avoids these boundary effects and so has a shallower gradient. This effect is not seen in the next $0.5<z_{max}\leq1$ kpc interval because the higher fraction of older thin disc stars and thick disc stars in this height interval flattens the gradient, reducing the boundary effect.

\section{Discussion and Conclusions}

We have investigated the Milky Way Galaxy's radial and vertical metallicity gradients using a sample of 47,406 red clump (RC) stars from the RAdial Velocity Experiment (RAVE) Data Release (DR) 4. The largest samples in the literature for investigating radial and vertical metallicity gradients are in \citet{Hayden14} and \citet{Boeche14}, who analyzed $\sim$20,000 stars, less than half of our sample.

Our sample was selected using the following constraints: (i) T$_{\rm eff}$ and log $g$ from the RAVE DR4 were used to select the RC stars, $1-\sigma$ from the peak density of the RC; (ii) the proper motions used in their space velocity estimations are available in the UCAC4 catalogue \citep{Zacharias13}, ii) the iron abundance is available in the RAVE DR4 catalogue \citep{Kordopatis13}; (iii) $S/N\geq40$; and (iv) total space velocity errors would be less or equal to 21 km s$^{-1}$. 

The mean absolute magnitude of the RC stars by \citet{Groenewegen08} ($M_{K_s}$=-1.54$\pm$0.04 mag) is used to determine distances to our sample stars. The resulting distances agree with the RAVE DR4 distances \citep{Binney14} of the same stars (mean and standard deviation of difference is 87 and 220 pc respectively). Our photometric method also provides distances to 6185 star (13\% of our sample) for which distances are not assigned in RAVE DR4.  

The metallicity gradients are calculated with their current orbital positions ($R_{gc}$ and $Z$) and with their orbital properties (mean Galactocentric distance, $R_{m}$ and $z_{max}$), as a function of the distance to the Galactic plane: d[Fe/H]/d$R_{gc}$=-0.047$\pm$0.003 dex kpc$^{-1}$ for $0\leq |Z|\leq0.5$ kpc and d[Fe/H]/d$R_m$=-0.025$\pm$0.002 dex kpc$^{-1}$ for $0\leq z_{max}\leq0.5$ kpc. This reaffirms the radial metallicity gradient in the thin disc but highlights that gradients are sensitive to the selection effects caused by the difference between $R_{gc}$ and $R_{m}$. The radial gradient is flat in the interval 0.5-1 kpc from the plane and then becomes positive for the distances greater than 1 kpc from the plane. Both the radial and vertical metallicity gradients are found to become shallower as the eccentricity of the sample increases.  

The radial metallicity gradients in terms of Galactocentric distance, for the interval of $0\leq|Z|\leq0.5$ kpc, d[Fe/H]/d$R_{g}$=-0.047$\pm$0.003 dex kpc$^{-1}$ is compatible with the one of \citet{Boeche14}, i.e. d[Fe/H]/d$R_{gc}$=-0.054$\pm$0.004 dex kpc$^{-1}$ within the errors, estimated for the RC stars with $5.5<R_{gc}<11$ kpc. The radial metallicity gradient in terms of mean orbital distance, d[Fe/H]/d$R_m$=-0.025$\pm$0.002 dex kpc$^{-1}$ is flat relative to the one of \citet{Bilir12}, d[Fe/H]/d$R_m$=-0.041$\pm$ 0.007 dex kpc$^{-1}$ estimated for the RC stars. However, \citet{Bilir12} restricted the vertical eccentricities of their star sample with $e_v\leq 0.07$. $e_v$ is proportional to $z_{max}$ and steeper gradients are found near  the plane.

The vertical metallicity gradient estimated for the present position of all RC stars, d[Fe/H]/d$|Z|$=-0.219$\pm$0.003 dex kpc$^{-1}$, is steeper than \citet{Boeche14}'s, d[Fe/H]/d$|Z|$=-0.112$\pm$0.007 dex kpc$^{-1}$. As the vertical metallicity gradient gets steeper further from the plane, our steeper result suggests we are sampling a higher number of stars further from the plane than \citet{Boeche14}.  

The vertical metallicity gradient in terms of the $z_{max}$ distance estimated in this study, d[Fe/H]/d$z_{max}$= -0.157$\pm$0.002 dex kpc$^{-1}$, is comparable with the one in \citet{Plevne15}, i.e. d[Fe/H]/d$z_{max}$= -0.176$\pm$0.039 dex kpc$^{-1}$. The difference between them may originate in the different $z_{max}$ intervals that the two gradients were evaluated, $0<z_{max}<3$ and $0<z_{max}\leq0.8$ kpc in our study and in \citet{Plevne15}, respectively. The d[Fe/H]/d$z_{max}$ metallicity gradients estimated for the RC stars in \citet{Bilir12} which are also given in Table 2 are different than the corresponding one in our study, due to additional constraints in \citet{Bilir12}.

Table 5 shows that the radial metallicity gradient depends on the eccentricity. Actually, the flat value of d[Fe/H]/d$R_m$= -0.025$\pm$0.002 dex kpc$^{-1}$ could be reduced to steeper values, -0.089$\pm$0.010, -0.072$\pm$0.007, -0.053$\pm$0.004 and -0.044$\pm$0.002 dex kpc$^{-1}$ by applying constraints $e_p\leq0.05$, 0.07, 0.10 and 0.20 in $0\leq z_{max}\leq0.5$ kpc distance interval, respectively. By this limitation, low $e_p$ steepen the gradients compared to including higher $e_p$. This could be due to many effects. $e_p$ is expected to increase with age and older stars tend to be more metal-poor so including higher $e_p$ stars includes more metal-poor stars, which will flatten the gradient. Similarly, thick disc stars tend to have higher $e_p$ than thin disc stars so including higher $e_p$ stars includes more thick disc stars, which also tend to be more metal-poor, again flattening the gradient.  

\begin{figure}
\centering
\includegraphics[width=9 cm,height=15.11cm]{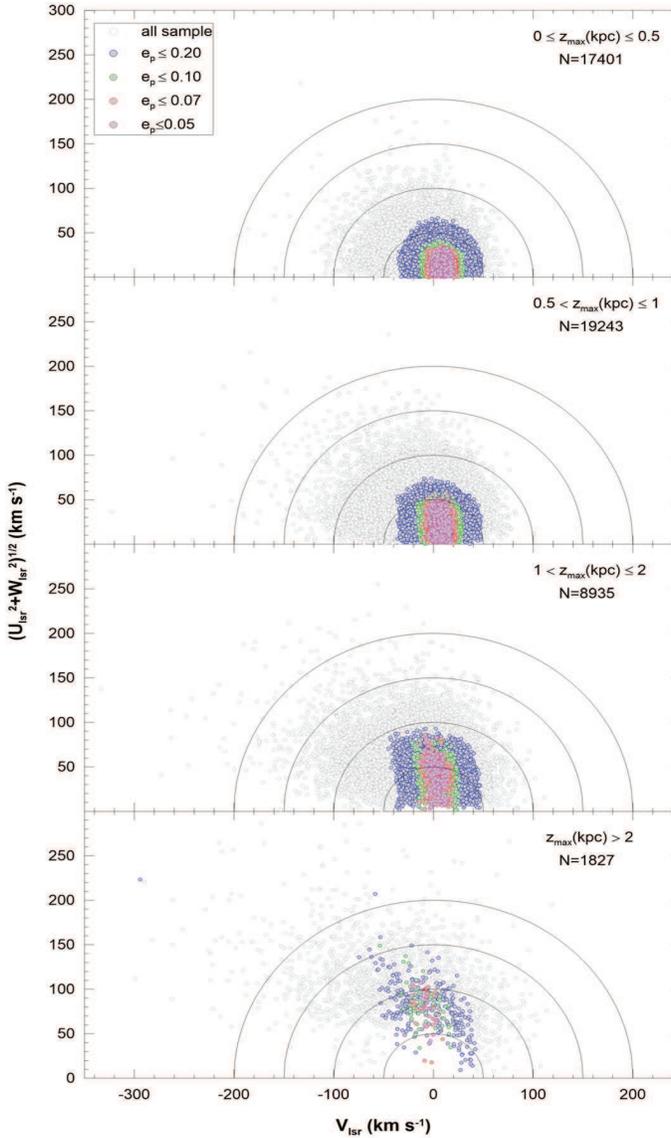}
\caption{Toomre energy diagram of 47,406 RC stars in four $z_{max}$ intervals. Pink, red, green, blue and grey circles represent $e_p\leq0.05$, $e_p\leq0.07$, $e_p\leq0.10$, $e_p\leq0.20$ and $e_p\leq1$ samples. Black solid lines show total space velocity borders of 50, 100 and 150 km s$^{-1}$, respectively.} 
\label{Fig16}
\end {figure}

We plotted the RC stars in the sub-samples defined by their $e_p$ eccentricities in four Toomre diagrams and investigated their behaviour in terms of space velocities. The Toomre diagrams in Fig. \ref{Fig16} cover the RC stars with distances $0\leq z_{max}\leq0.5$, $0.5<z_{max}\leq1$, $1<z_{max}\leq2$, and $z_{max}>2$ kpc. There is a trend between the space velocity components $U$ and $W$, and the $e_p$ eccentricities of the stars in all panels, i.e. $(U^2+W^2)^{1/2}$ increases with increasing $e_p$. Surprisingly, the total space velocity remains constant, which it suggests that the rotational velocity of the RC sample decreases with increasing $z_{max}$ distances. On the contrary, $(U^2+W^2)^{1/2}$ velocities of the stars for a given sub-sample increase with increasing $z_{max}$ distances. For example, $(U^2+W^2)^{1/2}\leq50$ km s$^{-1}$ for the stars with $e_p\geq0.20$ in the $0\leq z_{max}\leq0.5$ interval, while it is $(U^2+W^2)^{1/2}<100$ km s$^{-1}$ in the $1<z_{max}\leq2$ interval. Thus, we can say that the space velocity components of the RC stars at the same $z_{max}$ distance, but with different $e_p$ eccentricities, are different. The same argument holds for the stars with the same eccentricities with different $z_{max}$ distances. If we assume that different velocities originate from the intergalactic gas clouds of different angular momentum as well as different chemical structure, then we can expect different metallicity gradients for different sub-samples, as in case of this study.

These findings can be used to constrain different formation scenarios of the thick and thin discs. The next step will be to repeat this analysis with improved astrometry, including trigonometric parallaxes, measured by {\it Gaia}. {\it Gaia}'s first data release will include these measurements for Tycho-2 stars \citep{Michalik15}, which are the brighter half of RAVE stars.

\section{Acknowledgments}
Authors are grateful to the anonymous referee for his/her considerable
contributions to improve the paper. This study has been supported in part by the Scientific and Technological Research Council (T\"UB\.ITAK) 114F347 and the Research Fund of the University of Istanbul, Project Number: 36224. This research has made use of NASA's (National Aeronautics and Space Administration) Astrophysics Data System and the SIMBAD Astronomical Database, operated at CDS, Strasbourg, France and NASA/IPAC Infrared Science Archive, which is operated by the Jet Propulsion Laboratory, California Institute of Technology, under contract with the National Aeronautics and Space Administration.

\end{document}